\def\year{2018}\relax
\documentclass[letterpaper]{article} 
\usepackage[utf8]{inputenc}
\usepackage{aaai18}  
\usepackage{times}  
\usepackage[utf8]{inputenc}
\usepackage{blindtext}

\usepackage{helvet}  
\usepackage{courier}  
\usepackage{url}  
\usepackage{graphicx}  
\usepackage[inline]{enumitem}
\usepackage{subfig}
\usepackage{tikz}
\usepackage{placeins}
\usepackage[title]{appendix}
\expandafter\def\expandafter\UrlBreaks\expandafter{\UrlBreaks
  \do\a\do\b\do\c\do\d\do\e\do\f\do\g\do\h\do\i\do\j%
  \do\k\do\l\do\m\do\n\do\o\do\p\do\q\do\r\do\s\do\t%
  \do\u\do\v\do\w\do\x\do\y\do\z\do\A\do\B\do\C\do\D%
  \do\E\do\F\do\G\do\H\do\I\do\J\do\K\do\L\do\M\do\N%
  \do\O\do\P\do\Q\do\R\do\S\do\T\do\U\do\V\do\W\do\X%
  \do\Y\do\Z}

\usepackage[left=2cm, right=2cm, top=1.9cm, bottom=1.9cm]{geometry}
\usepackage[subtle]{savetrees}

\newcommand*{\affaddr}[1]{#1} 
\newcommand*{\affmark}[1][*]{\textsuperscript{#1}}

\begin{document}
%

\title{Analysis of Strategy and Spread of Russia-sponsored Content in the US in 2017}
\author{
    Alexander Spangher\affmark[1]\thanks{Research performed during an internship at Microsoft Research.}, Gireeja Ranade\affmark[2,3], Besmira Nushi\affmark[3], Adam Fourney\affmark[3], and Eric Horvitz\affmark[3] \\
    \affaddr{\affmark[1]Carnegie Mellon University}\\
    \affaddr{\affmark[2]University of California, Berkeley}\\
    \affaddr{\affmark[3]Microsoft Research}\\
}

\maketitle
\begin{abstract}
The Russia-based Internet Research Agency (IRA) carried out a broad information campaign in the U.S. before and after the 2016 presidential election. The organization created an expansive set of internet properties: web domains, Facebook pages, and Twitter bots, which received traffic via purchased Facebook ads, tweets, and search engines indexing their domains. We investigate the scope of IRA activities in 2017, joining data from Facebook and Twitter with logs from the Internet Explorer 11 and Edge browsers and the Bing.com search engine. The studies demonstrate both the ease with which malicious actors can harness social media and search engines for propaganda campaigns, and the ability to track and understand such activities by fusing content and activity resources from multiple internet services. We show how cross-platform analyses can provide an unprecedented lens on attempts to manipulate opinions and elections in democracies. 
\end{abstract}

\newcommand{\qq}[1]{``\textit{#1}''}

\noindent The Internet Research Agency (IRA) has been identified as a Russia-based company focused on media and information propagation \cite{ica2017}. The organization was found to have spent at least 5.8 million rubles, or sixty-eight thousand USD, from June 8, 2015 to July 1, 2017, on disseminating information to the U.S. public via Facebook advertising. They fielded thousands of Facebook advertisements and sent millions of Tweets, promoting hundreds of web domains and Facebook groups that spanned the political spectrum~\cite{schiff_facebook,schiff_twitter,clemson}. According to an indictment made against IRA by the U.S. Special Counsel's Office on February 16, 2018, the purpose of IRA's expenditures of effort and capital was to ``sow discord in the U.S. political system, including the 2016 U.S. presidential election''
\cite{mueller_ira_indictment}.
\begin{figure}[t]
    \includegraphics[width=1.1\columnwidth]{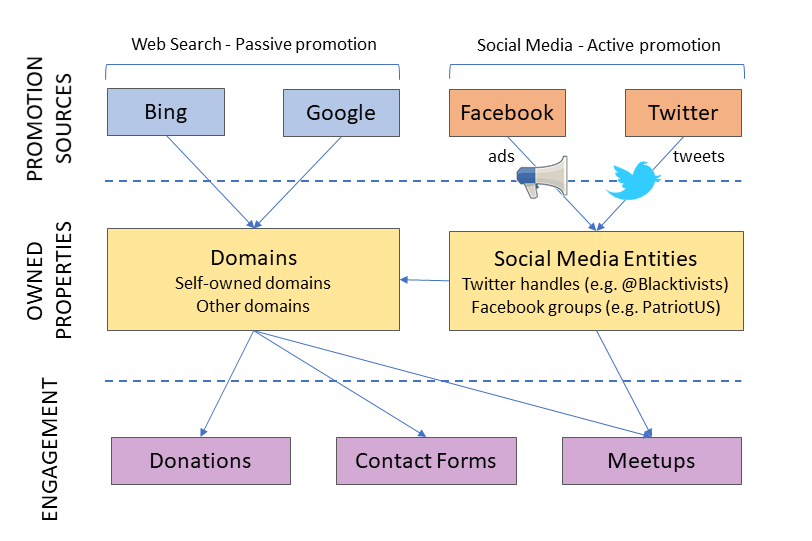}
    \caption{\textbf{IRA campaign structure.} Illustration of the structure of IRA sponsored content on the web. Content spread via a combination of paid promotions (Facebook ads), unpaid promotions (tweets and Facebook posts), and search referrals (organic search and recommendations). These pathways pointed to a combination of pages, some of which were created by the IRA (e.g. \url{blackmattersus.com}). IRA properties often contained engagement levers designed to maintain contact with users and push them towards events pages and donation pages. 
    }
    \label{Fig::GraphOverview}
\end{figure}

In the process of investigating Russian election interference, the U.S. House Intelligence Committee released IRA-linked Facebook advertisements and Twitter accounts into the public domain~\cite{schiff_facebook,schiff_twitter}. As stated in the release:
\begin{quote}
    Russia exploited real vulnerabilities that exist across online platforms and we must identify, expose, and defend ourselves against similar covert influence operations in the future. 
\end{quote}
We have pursued an understanding of \textit{how IRA's campaign targeted these ``vulnerabilities,'' in the hopes of contributing to an awareness of how malicious actors can exploit large-scale commercial internet services.} We merge Facebook and Twitter datasets with logs from Microsoft's web browsers and Bing search engine, thus fusing data from three major internet companies to provide a broader lens on the overall scope and influence of the IRA generated content (Fig.~\ref{Fig::GraphOverview}). 

We find that IRA-generated content spanned the
political spectrum and covered a combination of apolitical,
informative, local news and inflammatory articles on
politically sensitive subjects. During the time frame of our studies, we found that the IRA invested more resources on Facebook in promoting left-leaning
content than on right-leaning content. Right-leaning Twitter handles received more
traffic than left-leaning handles on Twitter.  
We describe two case studies that show that the apolitical content and local
news related IRA-properties reached users through search,
and played a role in bringing traffic to IRA domains. We examine the IRA Facebook ad keyword
targeting strategy and we investigate correlations between the regions where the ads received clicks, and the demographics of those regions.

\begin{figure}
    \centering
    \begin{tabular}{|c|}
    \hline
    \includegraphics[width=.39\textwidth, trim={.135cm 0 .05cm 0}, clip]{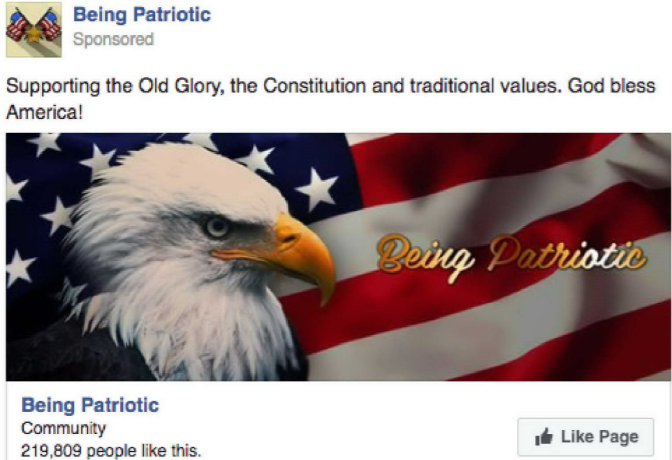} \\
    \hline
    \includegraphics[width=.403\textwidth, trim={.15cm 0 .25cm 0}, clip]{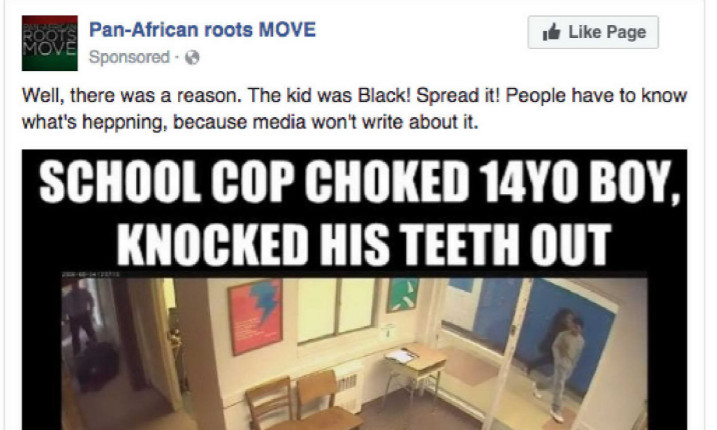}     \\
    \hline
    \end{tabular}
    \caption{Two example Facebook ads run by the IRA in 2017, out of 1,400 ads run post-election. The top ad is for the group ``PatriotUs'', while the bottom is for the group ``Pan-African roots''.}
    \label{fig:ExampleFacebookAds}
\end{figure}

Our specific findings are as follows:

\begin{enumerate}
    \item \textit{IRA Content:} The IRA produced and amplified a diverse array of left-leaning, right-leaning and apolitical content. We show two examples of Facebook advertisements in Fig.~\ref{fig:ExampleFacebookAds}. The advertisements and tweets ranged from emotionally charged to neutral (Fig. ~\ref{Fig::TrafficOverview}). IRA content drew traffic mainly from social media and search domains. While some IRA-established properties amplified politically sensitive topics (e.g. the Facebook group StopAllInvaders) others shared apolitical content\footnote{For example: "Black Female Computer Scientists. How Many Do You Know?" published at https://blackmattersus.com/31081-black-female-computer-scientists-how-many-do-you-know/}. One web property encouraged users to attend events, subscribe to mailing lists, and donate (see Tables ~\ref{table:promoted_content}, ~\ref{table:blmstructure},  Sections III, IV).
    \item \textit{IRA Strategies}:  
    Through a study of traffic to the IRA properties we are able to make educated guesses about strategies that the IRA might have employed. 
    For instance, we find that while there were similar amounts of 
    left and right-leaning tweets on Twitter, the right-leaning tweets received more traffic in our datatset (Fig.~\ref{Fig::TrafficOverview}).
    On the other hand, the IRA produced more left-leaning content on Facebook. Furthermore, they spent more money promoting left-leaning ads than right-leaning ads on Facebook.

We use a basic model to estimate the effect of the paid promotions on traffic spikes. Under certain assumptions and for our basic model we find the paid promotions likely increased traffic spikes to the left-leaning properties (Fig.~\ref{Fig::PromoEffects}).
        
        We identify a strategy of employing apolitical content to garner credibility and engagement; such content received traffic via Bing on Microsoft promotional channels. One IRA article about Black Female Computer Scientists\footnote{https://blackmattersus.com/31081-black-female-computer-scientists-how-many-do-you-know/} appeared among the search results returned for a query that was promoted across Microsoft device lock-screens (Fig.~\ref{Fig::BingImpressionChannels} and Section V).
    
    Finally, we present a case study on a tactic employed on Twitter: rapid tweeting about evolving local news stories (Fig. ~\ref{Fig::HenryBellos}). We believe a goal may have been to capture clicks before local news outlets were able to cover the story.
    
    \item \textit{IRA Outcome/Effects:} Despite the large volume of content generated by the IRA, on average only roughly $1$ in $40,000$ internet users in our dataset  clicked on an IRA-property (Tweet, Facebook Group, URL) on a given day. Furthermore, we found only low volumes of traffic to meetup pages, donation pages and contact forms on the IRA domains (Table \ref{table:blmstructure}). Likewise,  we found low correlation between geographic areas of heavy traffic to IRA domains and U.S. protests in 2017. We found no significant changes in traffic to news websites and to extreme news websites or to political donation websites (within a single browsing session, after a user was exposed to IRA content). This warrants further research that could take into account what we could not observe, such as traffic before and during the election, broader traffic from multiple browsers and search engines, traffic from mobile applications, long-term user behavioral change, real-world actions and posts made after exposure (see Section VI).
\end{enumerate}

The remainder of the work is structured as follows: We begin by describing our datasets and methodology (Section II). We then describe the high-level structure of the IRA web properties and promotions (Section III), and follow by discussing the themes and messages that were ultimately promoted and clicked on (Section IV). We discuss the impact of paid promotions on Facebook as well as the role played by search through some case studies (Section V). Finally we discuss the implications and limitations of this research, and contextualize it with prior research in this space.
 \section{II. Data and Methodology}
We have harnessed data from three large internet companies to provide a perspective on the strategy of the IRA, including views on the spread and access of IRA-generated content from January 2017 to August 2017. In addition, this work also leverages additional data sources including census information. We describe each data source in turn.

\subsection{Primary datasets}
\textbf{Facebook ads:} On May 10, 2018, the House of Representatives Permanent Select Committee on Intelligence released (as PDF documents) 3393 Facebook advertisements reported by Facebook as paid for by IRA-linked entities\footnote{PDF documents for the individual ads included: text associated with the ad, the ad's start and end dates, targeting information, and the URL of the web property that the ad was promoting.} \cite{schiff_facebook}. We performed optical character recognition (OCR) and template-based information extraction, and this process yielded 3061 ads with actionable data.

\noindent
\textbf{IRA-linked Tweets:} Alongside the Facebook ads data, the House Intelligence Committee also released a dataset on June 8th consisting of 3841 Twitter account handles believed to be associated with the IRA \cite{schiff_twitter}. Researchers at Clemson University scraped over 2.9 million tweets from these handles and performed language and topic classification \cite{clemson}. 

We leverage their work and focus our analysis on English-language tweets occurring from January 1st to August 1st, 2017 from the following categories they assign: ``LeftTrolls'', ``RightTrolls'', ``Local'' and ``News'' (See \cite{clemson} for more details). This resulted in a total of $471,000$ tweets and $320$ handles.

Additionally, many of the tweets link to external domains. We augment the data by resolving Twitter's link-shorteners using historic instrumentation logs from Microsoft web browsers, described next. 

\noindent
\textbf{Browsing data:} We consider 212 days (January 1 to August 1, 2017) of instrumentation data collected by Microsoft Edge and Internet Explorer 11 desktop web browsers.\footnote{Browser data is collected anonymously with user permission.} Data includes anonymized time-stamped records of page visits. Records are assembled into sessions of browsing activity: a ``session'' is defined as a sequence of page visits such that consecutive visits are less than 30 minutes apart. 

We also used this data to identify upload dates for photos and videos in Facebook groups. Facebook assigns photos and videos unique ids visible in the url (i.e. \url{facebook.com/<IRA group>/photos/<id>} and \url{facebook.com/<IRA group>/videos/<id>}). We group the content by these ids and recover the earliest dates for each ID that we observe clicks. This gives us a sense of when the item was posted. 

\noindent
\textbf{Web search data:} Finally, we also analyze the logs of Bing, a major web search engine. This data reveals which URLs were presented as search results, even if such pages were not ultimately clicked by users -- something that the browsing instrumentation data cannot provide. This dataset considers the same 212 days of data as the browser instrumentation logs.

\subsection{Joins of primary datasets}
We join the primary datasets to recover the scope of the IRA campaign. We associate IRA Facebook advertisements with browsing data by identifying clicks from \texttt{facebook.com} to a URL promoted by an IRA ad \footnote{After normalizing and reversing URL-transformations that Facebook applies to URLs (e.g. splicing \url{/pg/} or \url{/?/} from URLs). Such events are consistent with users clicking on ads, but are not sufficient to conclude that a particular ad was clicked because (a) a user may have arrived at a page from elsewhere in Facebook (e.g. from the newsfeed, or a notification) and (b) multiple ads promote a common URL.}. An equivalent approach was followed for joining browsing data with Twitter account handles and tweets \footnote{We searched for clicks with \url{twitter.com} and \url{<account handle>} as substrings, thus capturing clicks on tweets \textit{and} account handles.}.

Search engine logs were joined to the social media data by matching the URLs of search results to the URLs of pages actively promoted by Facebook ads and Tweets sent from IRA-linked twitter accounts. In search logs, we detect both when links are presented (impressions) and clicked. 

\subsection{Crowdsourced labeling of primary data}
We used crowdworkers to label content\footnote{We use Amazon Mechanical Turk. Workers were paid on average \$12 per hour, above the national minimum wage at the time of publication.} to ascertain information about the IRA content's political leaning, emotional intensity, and topic. We ran this study for Facebook ads, tweets, and search results related to IRA-owned web domains.

For Facebook ads, we rated all 1032 ads that ran after January 1, 2017. For Twitter we rated two subsets: a random subset of 500 IRA-linked tweets, and the top 500 most clicked tweets. For Search, we rated the 100 URLs with the most impressions (these 100 URLs captured 63\% of search-result impressions to URLs in our datasets).

The task showed workers the text content of the promotion: the original Facebook ad, tweet text, or the snippet text shown in search results. Based on this information, the workers' task was to label the content with the most relevant option in each of the following categories:
\begin{enumerate*}
    \item \emph{Political leaning}:~\{extreme-left, left, center, right, extreme-right, apolitical, local news\}. 
    \item \emph{Emotional intensity}:~\{neutral, low, medium, high, very high\}.
    \item \emph{Discussed topic}:~A predefined topic list.\footnote{To extract the topic list we grouped Facebook ad keywords based on their co-occurrence across advertisements. These were then manually labeled (e.g. Black Lives Matter, Veterans, see Figure ~\ref{fig:TopicsContent} for a partial list.) See Section IV for more discussion on targeting tags.}
\end{enumerate*}

Each item was judged by five different crowd workers. We did quality control by excluding answers from workers with high disagreement rates~\cite{inel2014crowdtruth}. We performed a soft-assignment to label each item. For example, if a certain URL received 2 votes for extreme-right, 1 for left and 2 for extreme-left, then we counted $2/5, 1/5$ and $2/5$ clicks in each category respectively. To calculate the number of clicks on a certain label/topic, we used the soft assignment to proportionally divide the clicks across the assigned labels. 

\subsection{Secondary datasets}
In addition to the primary datasets, we employ numerous external secondary datasets. We use Mediabias News and Media Categories \cite{mediabias} and SimilarWeb Domain Categories \cite{similarweb}, which provide URL-level categorizations, to characterize the content of URLs in our datasets. We use state-level popular vote counts for the 2016 presidential election \cite{votes2016}, state-level voter registration data \cite{voterregistration2016}, state-level census demographic data \cite{census2010}, and state-level protest counts over time \cite{crowdcounting} to characterize browsing data by geography. GDELT Global-Events data, which captures media publications over time, is used to characterize external news events \cite{gdelt}. Finally, we use OpenSecrets.org \cite{opensecrets} and Ballotpedia \cite{ballotpedia} to identify major political and donation sites over the period of interest.

 \section{III. Structure of the IRA Promotions}
Before describing the traffic outcomes of the IRA campaign in Section IV, we present an overview of the scope and structure of the campaign to illustrate the breadth of the IRA's activity on different platforms.

\begin{table}
    \centering
    \footnotesize
    \begin{tabular}{|p{3.5cm}|r|r|}
    \hline
        \textbf{Promoted Content} & \textbf{\# Properties} & \textbf{\# Ads} \\
        \hline
        \hline
        Facebook groups & 104 & 2674\\
        Events \& meetups & 82 & 223\\
        IRA domains & 4 & 128 \\
        \hline
        News organizations & 4 & 4\\
        Other & 16 & 35\\ 
        \hline
        \# Total & 207 & 3061 \\
        \hline
    \end{tabular}
    \caption{\textbf{Facebook Ad Categories.} Web properties promoted by the IRA Facebook campaign includes Facebook groups, events and domains suspected to be under the editorial control of the IRA~\cite{schiff_facebook}.}
    \label{table:promoted_content}
\end{table}

\subsection{Facebook Ads}
We found that the Facebook advertisements paid for by the IRA promoted a smaller set of $350$ distinct URLs. These URLs were distributed across 207 web properties. These properties include Facebook groups and profiles; Facebook or \url{meetup.com} event pages; news websites (including \url{CNN.com}); petitions (\url{whitehouse.gov}, \url{change.org}); and four domains identified as controlled by the IRA (\url{blackmattersus.com}, \url{dudeers.com}, \url{black4black.com}, \url{donotshoot.us})\footnote{See https://intelligence.senate.gov/sites/default/files/documents/exhibits-080118.pdf, https://euronews.com/2018/05/10/sean-hannity-black-lives-matter-among-targets-russian-influence-campaign-n872926, https://money.cnn.com/2017/10/12/media/dont-shoot-us-russia-pokemon-go/.} (See Table \ref{table:promoted_content}). 

The breadth of different content-types is notable (petitions, events, articles, Facebook groups), and raises questions about the intent of the IRA campaign.  In the next section covering traffic patterns, we limit our analysis to traffic. 
However, we note that the reach of the IRA campaign did go beyond what can be measured by just traffic. For instance, one of the IRA's petitions\footnote{https://www.change.org/p/barack-obama-u-s-house-of-representatives-u-s-senate-list-the-ku-klux-klan-as-an-official-terrorist-organization} received 65,000 supporters, well within the range of typical mid-to-high performing \url{change.org} campaigns (according to top popular petitions listed at \url{https://www.change.org/petitions}).

\subsection{Twitter}
The $3,841$ Twitter handles released by \cite{schiff_twitter} and 2.9 million tweets compiled by \cite{clemson} were filtered to $471,000$ English-language tweets from $320$ ``LeftTrolls'', ``RightTrolls'', ``Local'' and ``News'' accounts. Our analysis of links in these tweets revealed links to over $5,500$ domains, tweets, and other properties.
 Over $95\%$ of tweets are retweets, or reposting of other users' tweets, and many of the tweets point to well-known websites.

We found overlap in naming between the Facebook and Twitter campaigns: for instance, the Twitter campaign included a Twitter handle ``blackmattersussoldier'' and the Facebook campaign included a ``BlackMattersUS'' Facebook Group. However, we do not see the Facebook and Twitter campaigns sharing links or cross-promotions with each other. We observed only four URLs that were promoted both by Facebook ads and Tweets, all of them \url{blackmattersus.com} URLs with low-traffic.

We could only identify a small number of domains from the Twitter campaign as IRA-controlled (as compared to the Facebook campaign), and many of the links promoted by IRA tweets included major domains such as \url{nytimes.com}.
As a result we focus on an analysis of the tweets themselves (for the subset outlined in Section II), and not the domains they promoted.

\subsection{Web Search}
According to the February 2018 indictment by the Special Counsel's office, the IRA was \qq{organized into departments, including: a graphics department; a data analysis department; a search-engine optimization (SEO) department; an information-technology (IT)...} \cite{mueller_ira_indictment}. We find that IRA-promoted web properties were indexed by search engines, and surfaced to users in response to organic web search queries. Of the suspected IRA-owned domains we find one domain in particular, \url{blackmattersus.com}, received significant traffic via Bing (Fig. ~\ref{fig:SearchSocialEmotions} and Fig. ~\ref{Fig::BingImpressionChannels}). We did not find evidence of paid advertising on Bing, but more research is needed for a comprehensive evaluation.

\subsection{Targeting Decisions}
The IRA used 890 different targeting keyword across the Facebook ads. The most frequent used keywords are all politically charged or demographically salient, with the top five being ``african american civil rights movement'', ``african american history'', ``malcolm x'' and ``martin luther king jr''. These five keywords are tagged to 32.3\% of ads, which link to pages capturing 28\% of traffic.

We identified pairs of keywords used together with correlation  $c>.6$, ($p< .01$), then grouped these pairs together to identify clusters. We identified 19 such clusters using this method. When used together the keywords seem to indicate an attempt at demographic targeting.
(We will analyze the demographic reach of the IRA campaign in Section V).

On Twitter, the most frequently used hashtags (of 44,700 total unique hashtags) are a combination of political and apolitical tags,  with the top five being: ``\#MAGA\footnote{Make America Great Again}'', ``\#NowPlaying'', ``\#tcot\footnote{Top conservatives on Twitter}'', ``\#top'' and ``\#PJNET\footnote{Patriot Journalist Network}''. The top ten are tagged to 2.1\% of tweets which themselves account for 1.3\% of traffic: hashtags on Twitter were more diverse and did not repeat as frequently as the Facebook ads keywords.

\section{IV. IRA Content and Traffic}
Having described the breadth and size of the IRA's campaign in Section III, we now summarize the content of the campaign and provide an overview of the traffic to it. 

\subsection{Traffic Overview}
As noted in the introduction, on average roughly $1$ in $40,000$ internet users was exposed to IRA ads on any given day in our dataset.

Perhaps because the platforms' audiences are different \cite{facebook_twitter_audiences}, or because there was little cross-promotion between the Twitter and Facebook IRA campaigns, we find very small overlap between users exposed to both campaigns. Cross-traffic between IRA Facebook properties and IRA tweets is also small, amounting to about 0.02\% of the total users in our datasets clicking on content from both campaigns in the same day.

Table \ref{table:topTraffic} shows the most trafficked Facebook groups and Twitter handles in each campaign, which account for more than $3/4$th of total observed traffic in each case. 
The top-clicked Facebook groups seem to show a mix of topics and political leanings. 

\begin{figure}[t]
    \centering
    \footnotesize
    \begin{tabular}{ll}
    \subfloat[Referral pathways for clicks on URLs included in the IRA Facebook ads.]{
    \includegraphics[width=.21\textwidth, height=.11\textheight]{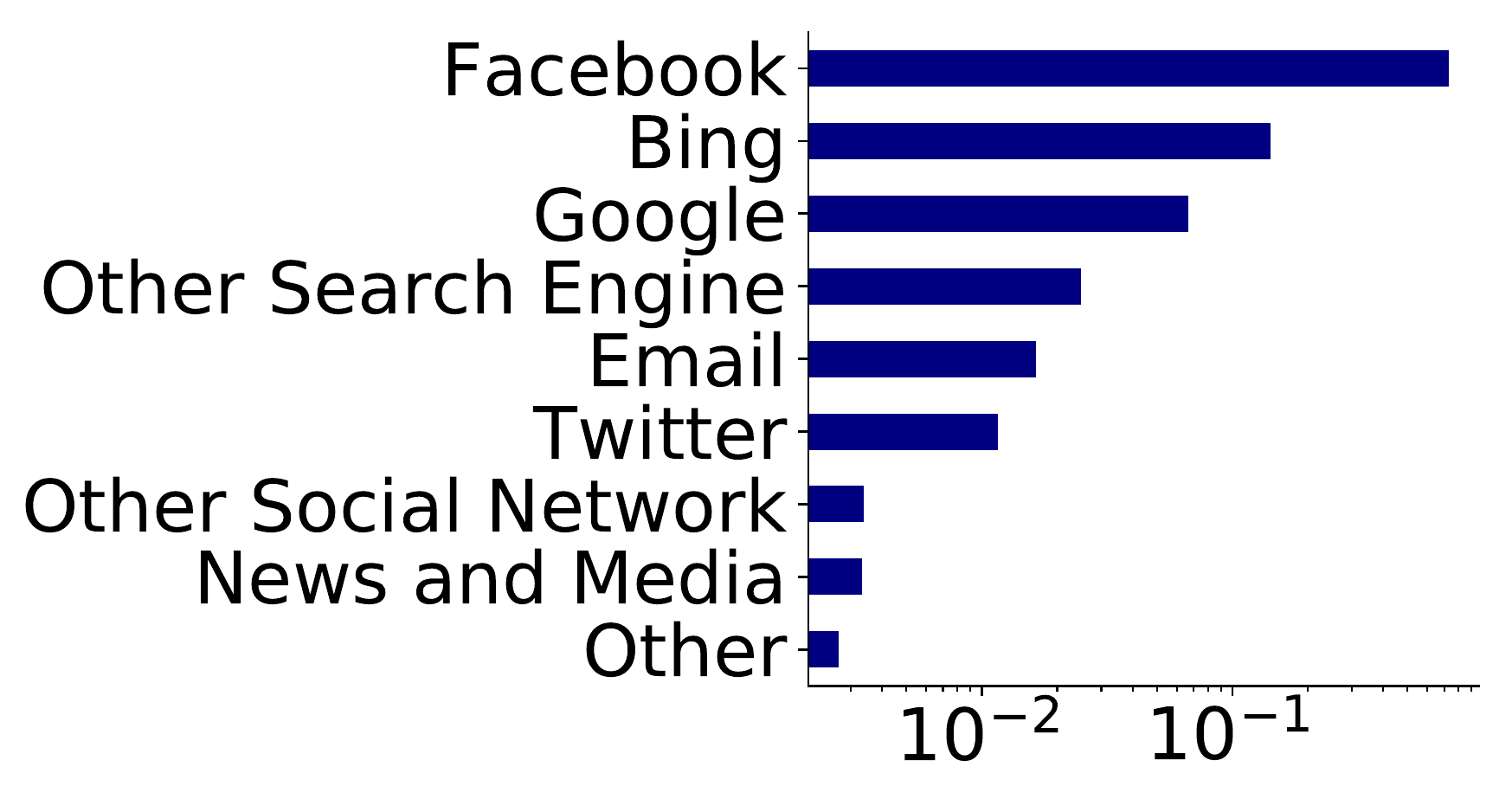} 
    \label{fig:refFacebook}
    } \hspace{2mm}
    \subfloat[Referral pathways for clicks on tweets linked to IRA accounts.     ]{\includegraphics[width=.21\textwidth,height=.11\textheight]{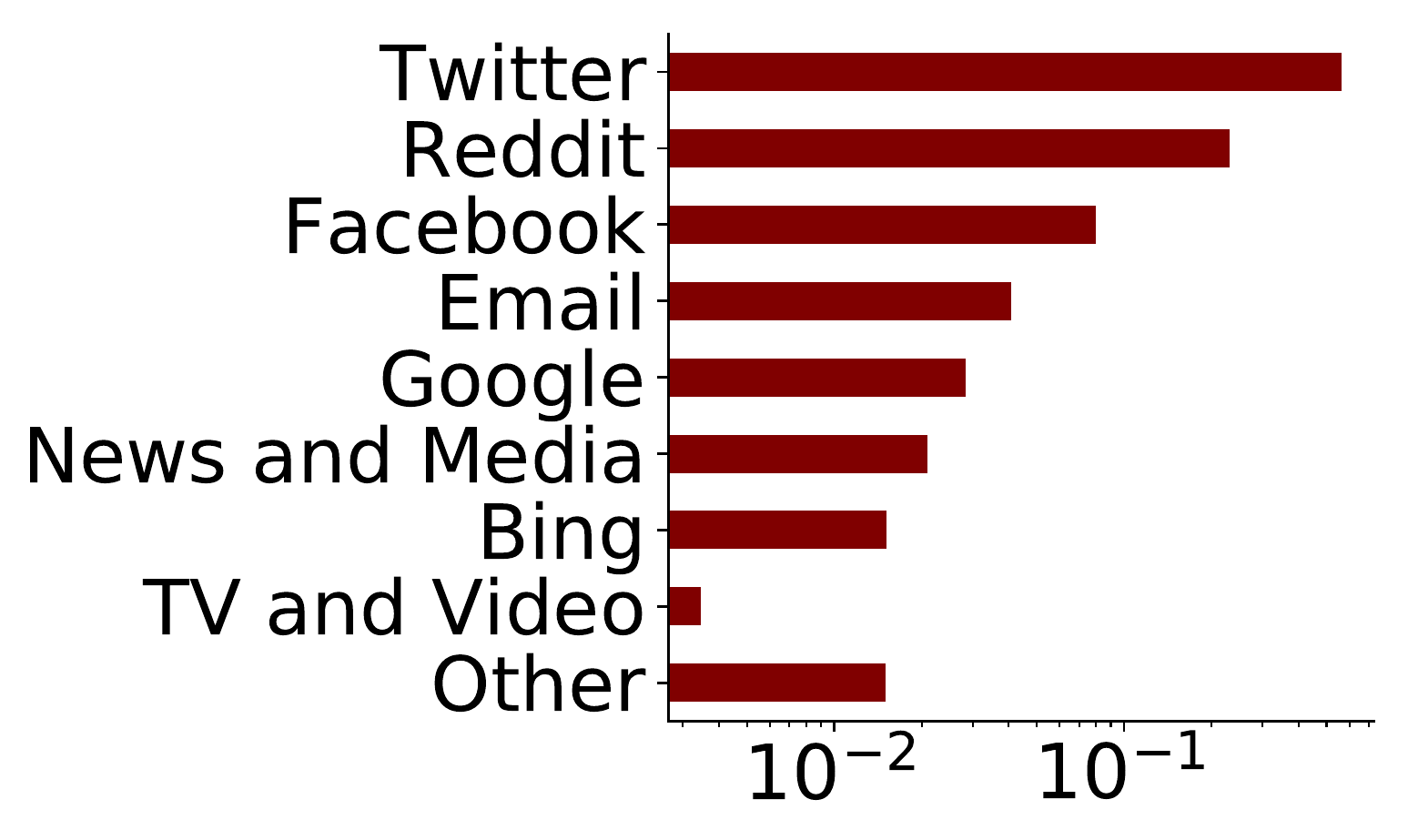}
    \label{fig:refTwitter}
    }
    \end{tabular}
    \caption{\textbf{Top traffic channels to IRA properties.} Top sources that brought users to IRA properties (Facebook-advertised URLs, Facebook groups and Tweets) from January 1, 2017 to August 1, 2017.}
\label{Fig::TrafficOverview}
\end{figure}

Figures ~\ref{fig:refFacebook} and~\ref{fig:refTwitter} summarize the common referral pathways to IRA properties.
Traffic to Facebook-advertised URLs and groups came mainly from Facebook and Search.  
The Twitter campaign drew traffic mainly from Twitter, Reddit and Facebook, with Search playing a much smaller role. 

\begin{table}
\centering
\footnotesize
\begin{tabular}{|l|r|}
\hline
\textbf{Top-Clicked Facebook Groups} &  \textbf{Share of IRA Traffic} \\
\hline
blacktivists         &          0.208 \\
godblessthesouth     &          0.199 \\
blackmattersus.mvmnt &          0.169 \\
brownunitedfront     &          0.116 \\
patriototus          &          0.077 \\
Other                             &          0.231 \\
\hline
\textbf{Top-Clicked Twitter Handles} &  \textbf{Share of IRA Traffic}\\
\hline
ten\_gop         &          0.462 \\
pamela\_moore13  &          0.245 \\
crystal1johnson &          0.076 \\
southlonestar   &          0.048 \\
jenn\_abrams     &          0.045 \\
Other           &          0.123 \\
\hline
\end{tabular}
\caption{\textbf{Top-performing IRA properties.} The top five Facebook groups (top) and Twitter handles (bottom) with most traffic. These Facebook groups collectively account for 77\% of traffic to IRA-promoted Facebook properties, while these Twitter handles account for 88\% of traffic.}
\label{table:topTraffic}
\end{table}

\subsection{Traffic by Content-Type}
\begin{figure}[t]
\centering
\begin{tabular}{@{}l|l@{}}
\subfloat[Distribution of political leaning of IRA Facebook posts and tweets. We see more left-leaning and apolitical links than right leaning links on Facebook. Twitter has roughly equal amounts of right and left content.]
{\includegraphics[width = .22\textwidth,height=.112\textheight ]{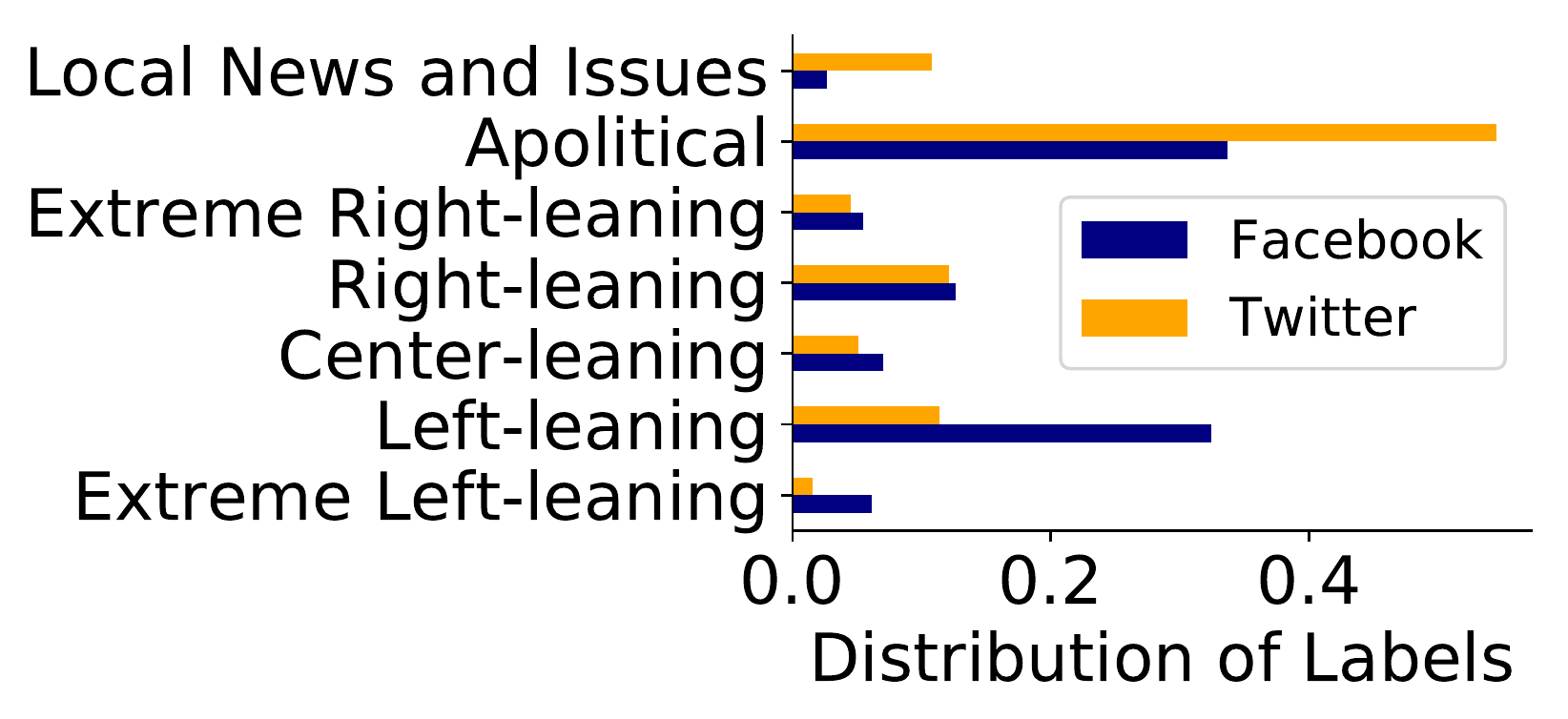} \label{fig:PoliticsContent}
}
& 
\subfloat[
Distribution of traffic to different categories of IRA Facebook posts and tweets, by political-leaning.  Even though there was more left-leaning content on Facebook than right-leaning, the right-leaning content took a larger share of the traffic.]
{\includegraphics[width = .22\textwidth,height=.112\textheight ]{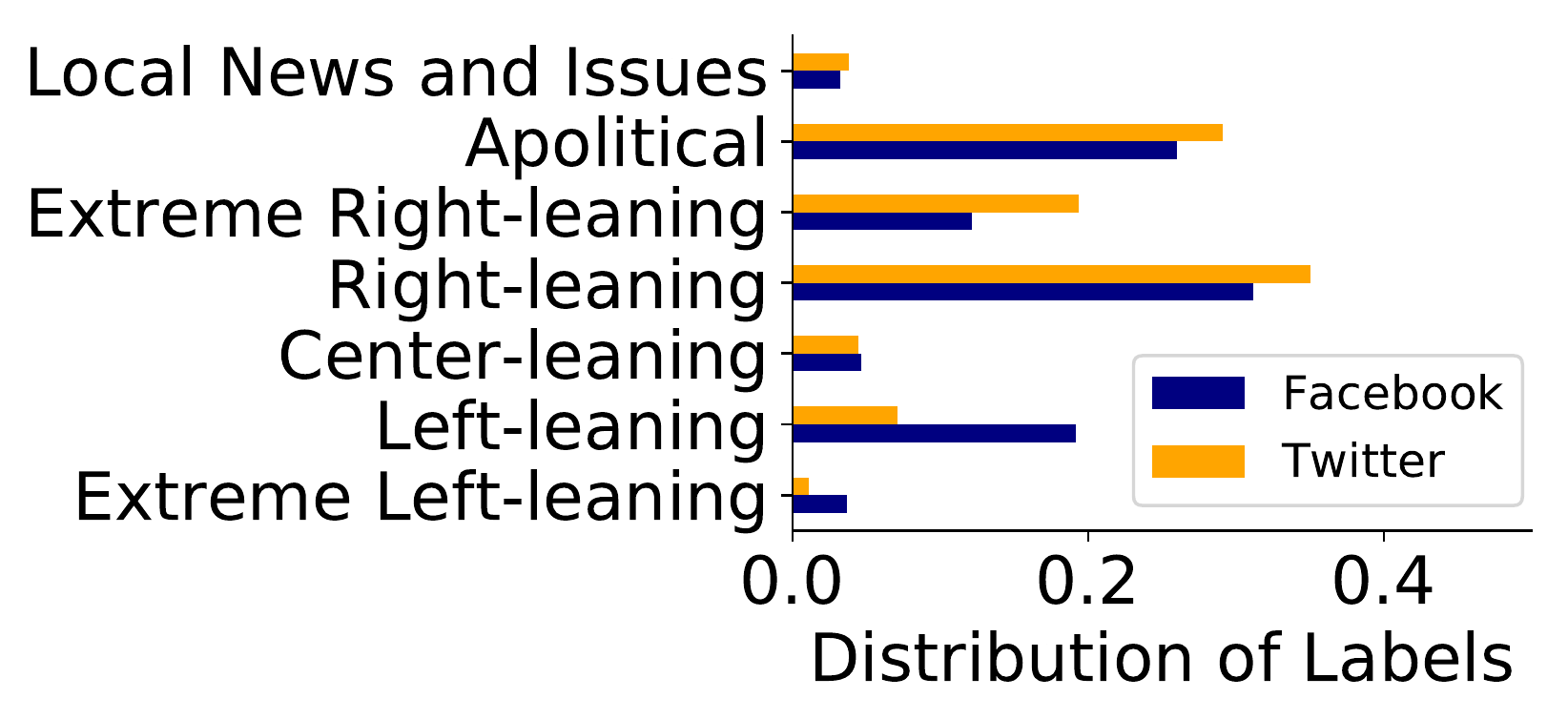}
\label{fig:PoliticsTraffic}
}\\

\subfloat[][Distribution of content across emotional intensity tags. A large fraction of the IRA promoted Facebook and Twitter content was neural and low emotional intensity. Especially pronounced is the Twitter channel, where nearly 43\% of content is rated "Neutral".]{
\includegraphics[width = .22\textwidth, height=.12\textheight]{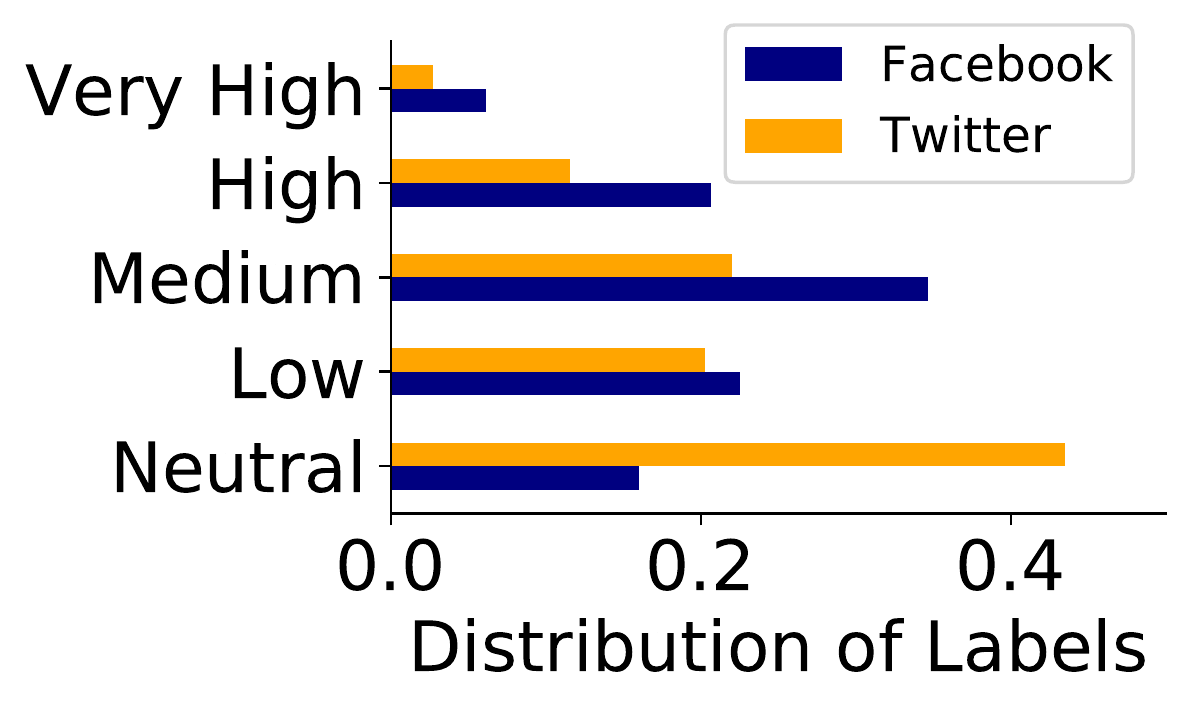}\label{EmotionsContent}
} &
\subfloat[][Distribution of traffic across the emotional valence of content. Relative to the proportions of overall content produced \protect\subref{EmotionsContent}, content with medium and high emotional valence received more traffic on Twitter.]{
\includegraphics[width = .22\textwidth,height=.12\textheight]{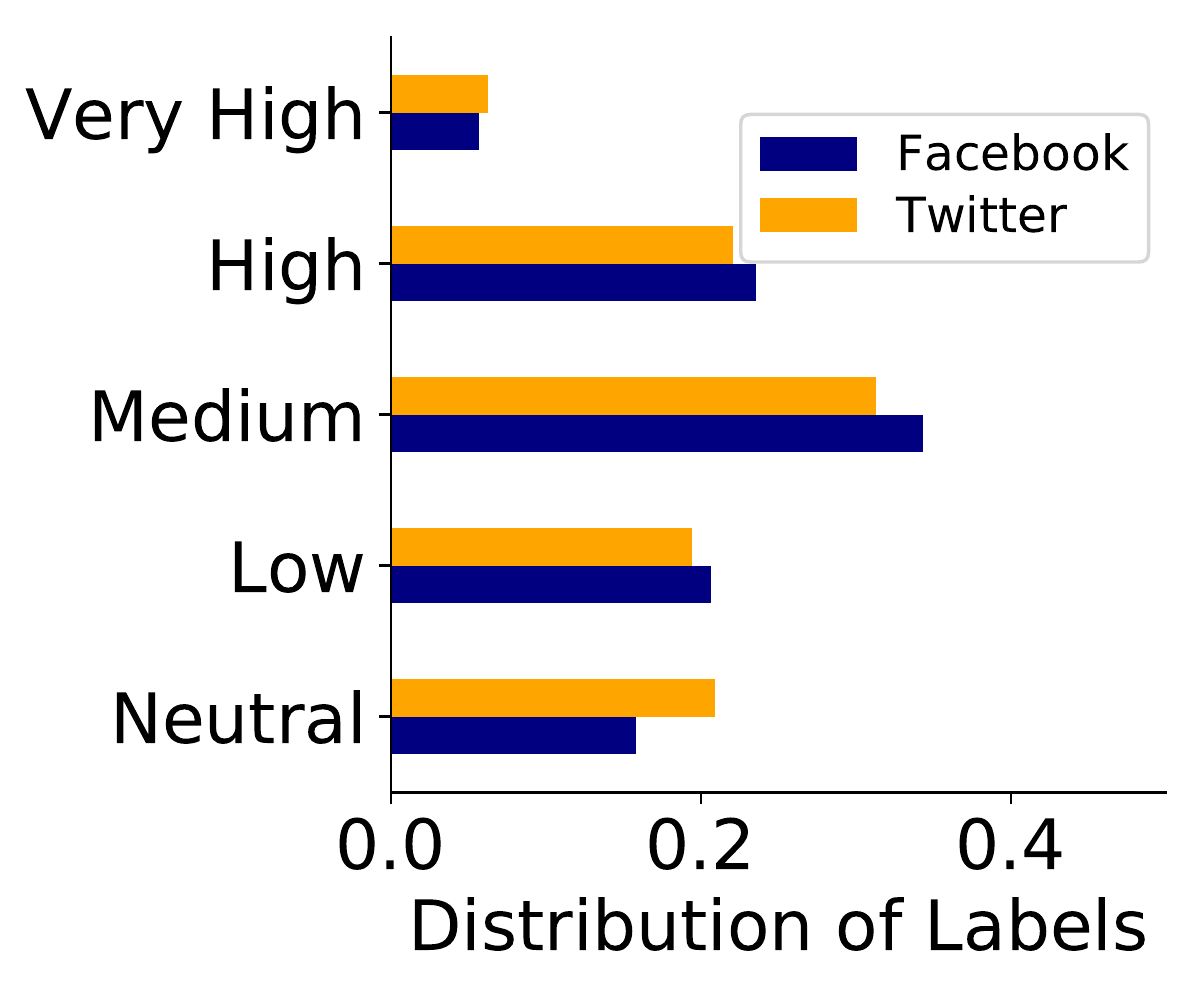}\label{EmotionsTraffic}
} \\ 

\subfloat[Distribution of IRA content across topics (Facebook and Twitter).
]{
\includegraphics[width=.22\textwidth, height=.2\textheight]{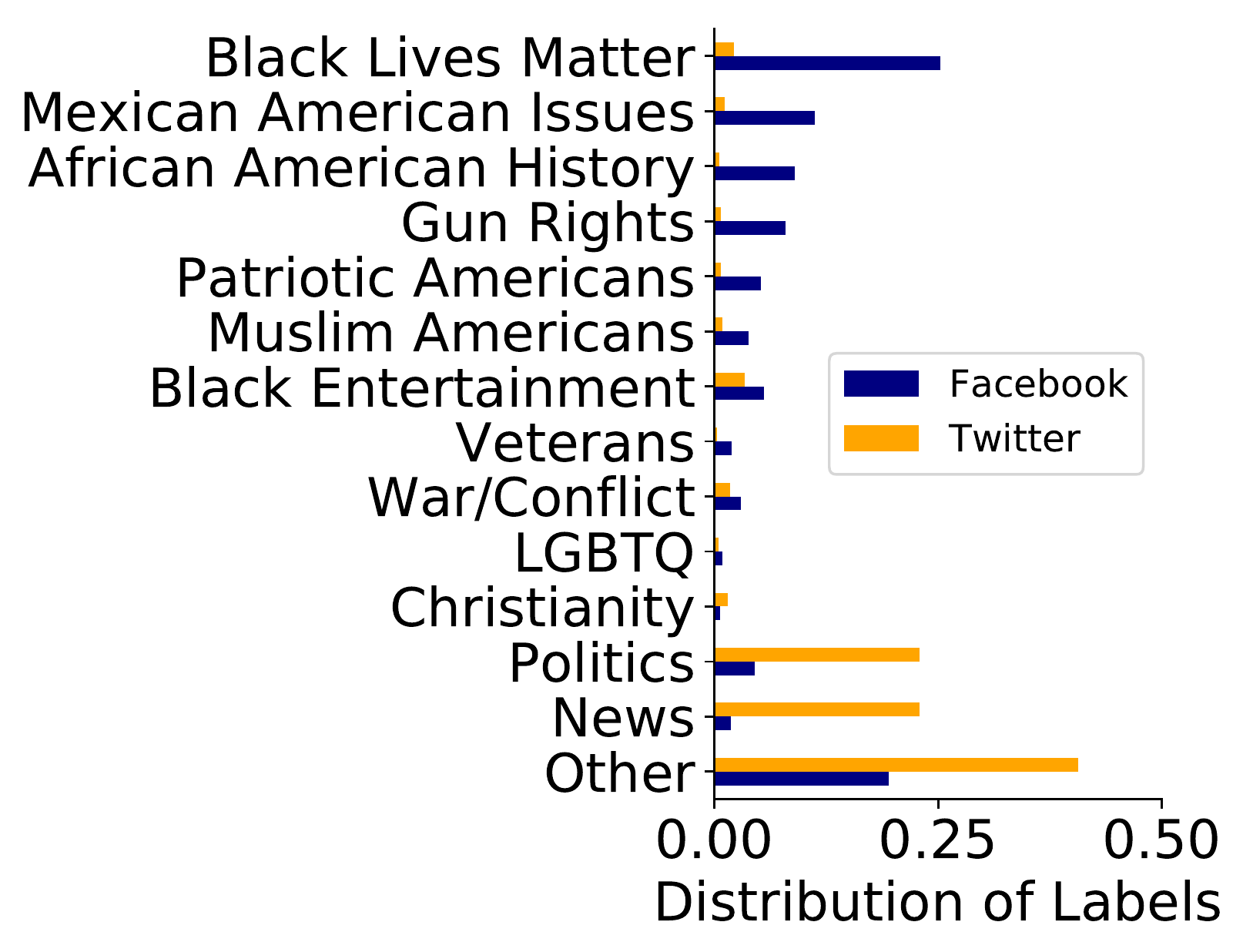}
\label{fig:TopicsContent}
} 
&
\subfloat[Distribution of traffic to IRA content across topics (Facebook and Twitter).
]{
\includegraphics[width=.22\textwidth, height=.2\textheight]{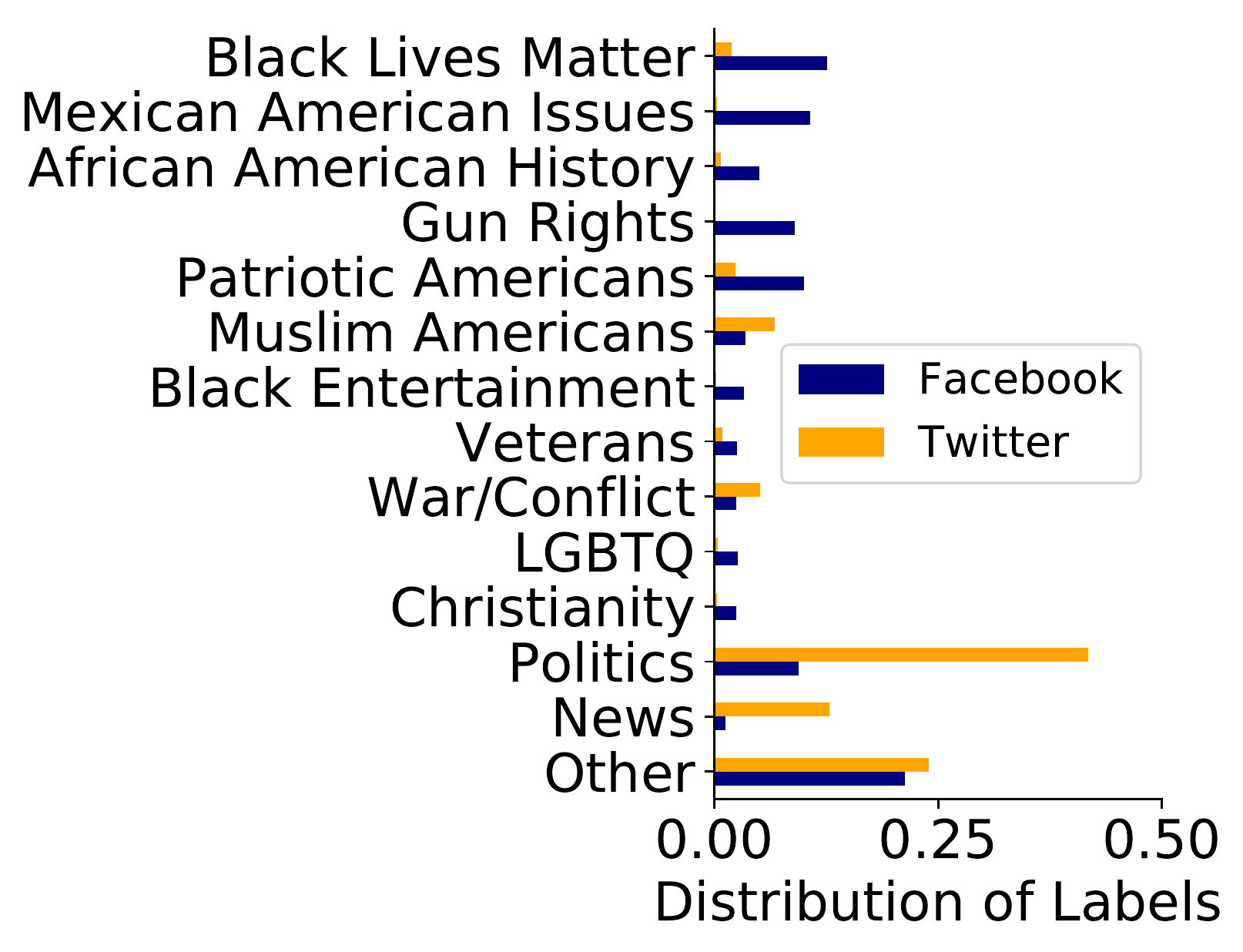}
\label{fig:TopicsTraffic}
}
\end{tabular}
\caption{\textbf{Labels Summary: Content and Traffic.} This figure shows the distribution of the IRA content across political leaning, emotional intensity and specific politically charged topics using labels tagged in a crowdsourcing experiment. The assignment of content to categories was done using the soft-assignment of crowdsourcing labels as described in the methodology section.
}
\label{fig:mturk_summary}
\end{figure}

Figure~\ref{fig:mturk_summary} summarizes results from our crowdsourcing study for both Facebook and Twitter. The left column summarizes the IRA generated content. The right column describes the traffic (volume of user cliks) to this content. The figure was created using the proportional assignment of clicks as described in the methodology section.  
The X-axis in all the histograms shows the distribution of content (ads or tweets) soft-assigned to each category on the Y-axis.

We see from Fig.~\ref{fig:PoliticsContent} that there were more left-leaning and apolitical ads than right-leaning ads on Facebook. The Twitter content appears to be more balanced, with most tweets being apolitical and roughly equal numbers of left-leaning and right-leaning tweets. When we compare this to the traffic that was received in each category, we see that right-leaning and apolitical content got relatively more traffic in our dataset than left-leaning content on both Facebook and Twitter. 

In addition to covering both sides of the political spectrum, the IRA content spanned a large topic space: Fig.~\ref{fig:TopicsContent} gives the distribution of content (ads, snippets, tweets) across categories, and Fig.~\ref{fig:TopicsTraffic} gives the distribution of traffic\footnote{Topic labels were chosen by manual assignment based on clusters among targeting tags in the Facebook ads. The assignment of ad/tweet to topic was done by crowdsourcing.}. Especially prominent in the Facebook campaign were topics targeting African American and Mexican American interests. The Twitter campaign seems to have focused on general news, politics and other topics.

 \section{V. IRA Strategy}
Having summarized the IRA content and traffic to it in Section III, we now share inferences about IRA's strategies. We start with three case studies.
\begin{itemize}
    \item \textbf{Case study 1:} Rapid-response tweeting to local news developments helped the IRA receive search engine impressions.
    \item \textbf{Case study 2:} 
    Apolitical IRA-generated stories appeared among search results returned for search queries on Bing.com, including some Microsoft-promoted queries (e.g., ``women scientists''). Overall, apolitical articles from organic and Microsoft-promoted queries received more traffic via the Bing.com search engine than other articles.
    \item \textbf{Case study 3:} Paid promotions were not the only traffic driving factor. In fact, we observe that for some of the groups, there exist other important unpaid actions (e.g. photo and video shares) that can predict high traffic.
\end{itemize}

We conclude this section by showing how the IRA reached various demographically and politically distinct geographic regions in the country and extensively used Facebook microtargeting tags.

\subsection{Case study: Search and \url{twitter.com/TEN_GOP}}
The IRA tweeted rapidly and in huge volumes across their accounts, tweeting in some cases thousands of times a day, and multiple times within seconds. Figure ~\ref{fig:mturk_summary} shows that they principally tweeted about ``News'', ``Politics'', and ``Entertainment''. The high volume of rapid tweets allowed them to be unwittingly incorporated into legitimate news articles at major Western outlets (For example, an article published by \url{msn.com}\footnote{https://www.msn.com/en-gb/news/uknews/london-attack-woman-wearing-hijab-was-distressed-horrified-photographer-says/ar-BByFoiY} references a tweet by @SouthLoneStar, a confirmed IRA account).

In this example, we show that these tweets performed especially well at drawing unwitting traffic from users discovering a new news story. The IRA's TEN\_GOP\footnote{TEN\_GOP refers to Tennessee GOP. Many IRA accounts mimicked political accounts~\cite{mueller_ira_indictment}.} Twitter handle, which drew 46\% of traffic in the IRA's Twitter campaign, is a good case study of how an IRA Twitter account can capture significant traffic by being indexed by a search engine during the very early stages of a breaking news cycle.

\begin{figure}
\subfloat[Spike in search impressions (normalized as percent of max) that displayed the TENGOP tweet about Henry Bello on June 30, 2017.]{
\includegraphics[width=.45\textwidth]{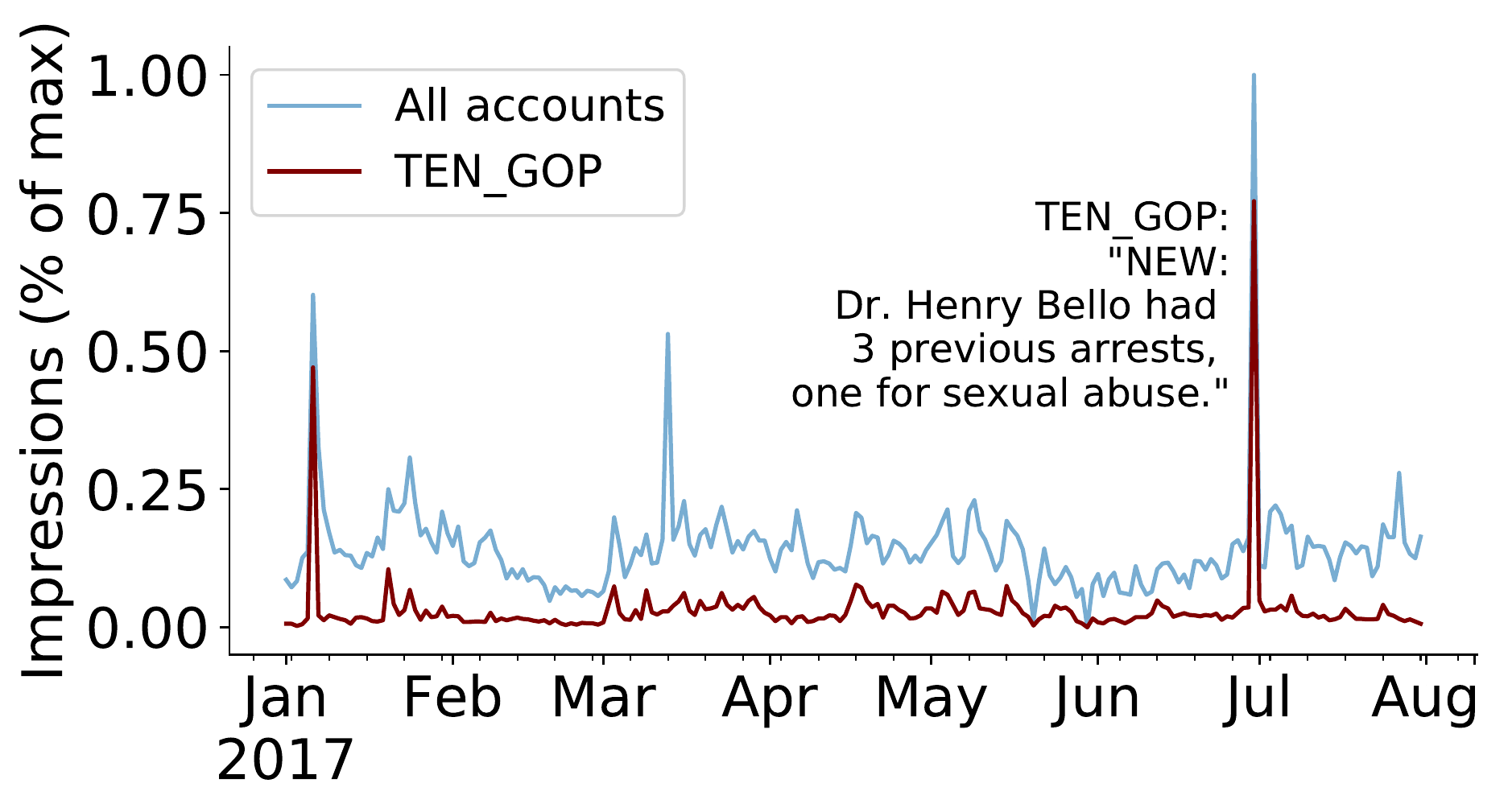}
}
\\
\subfloat[Distribution over impressions received by the TENGOP handle, compared with the number of news articles published on the subject on each day.]{    
    \includegraphics[width=.215\textwidth, height=.12\textheight]{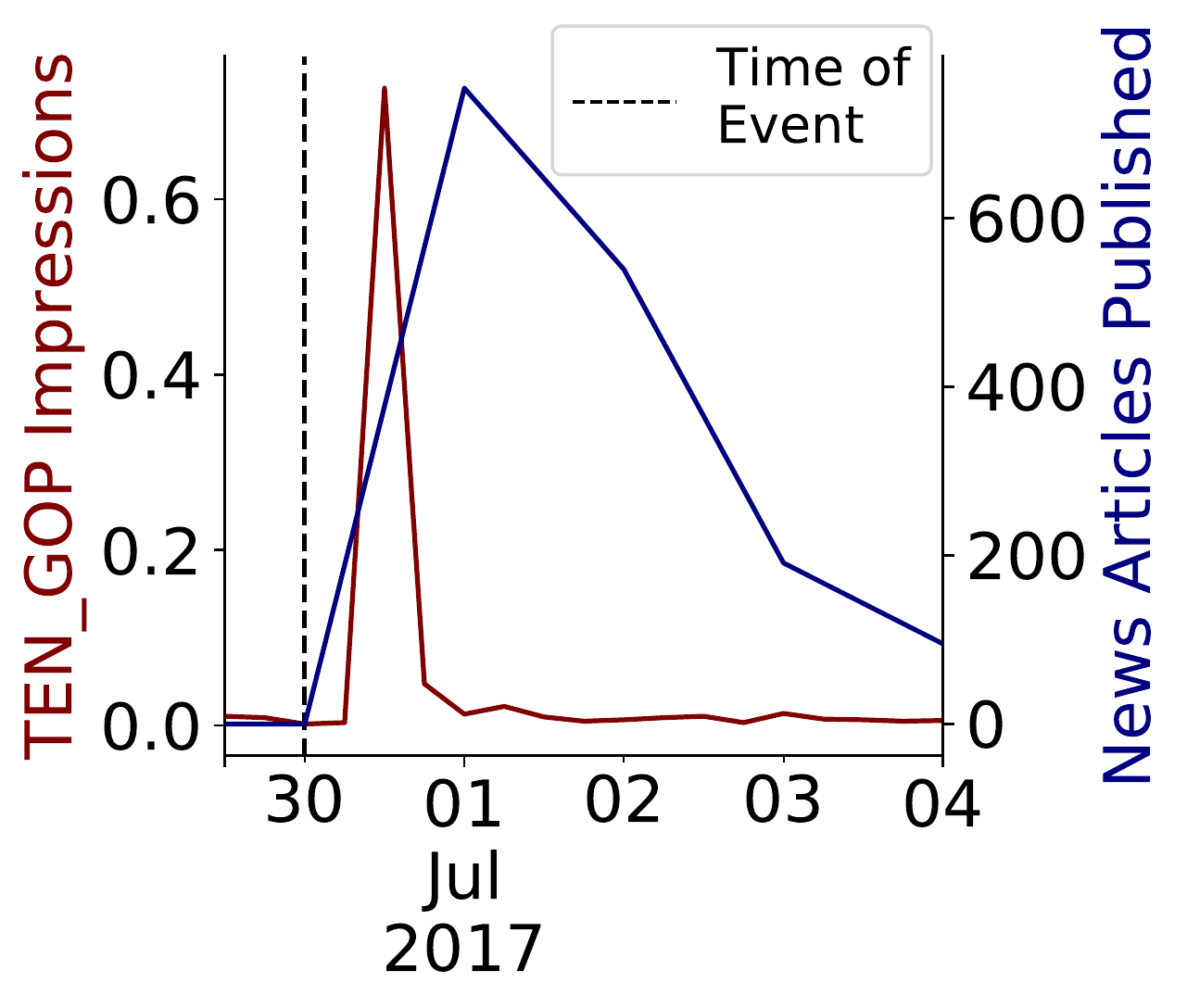} 
    } \hspace{.5mm}
    \subfloat[Top search terms leading to impressions received in subfigure (a).]{
    \includegraphics[width=.215\textwidth, height=.12\textheight]{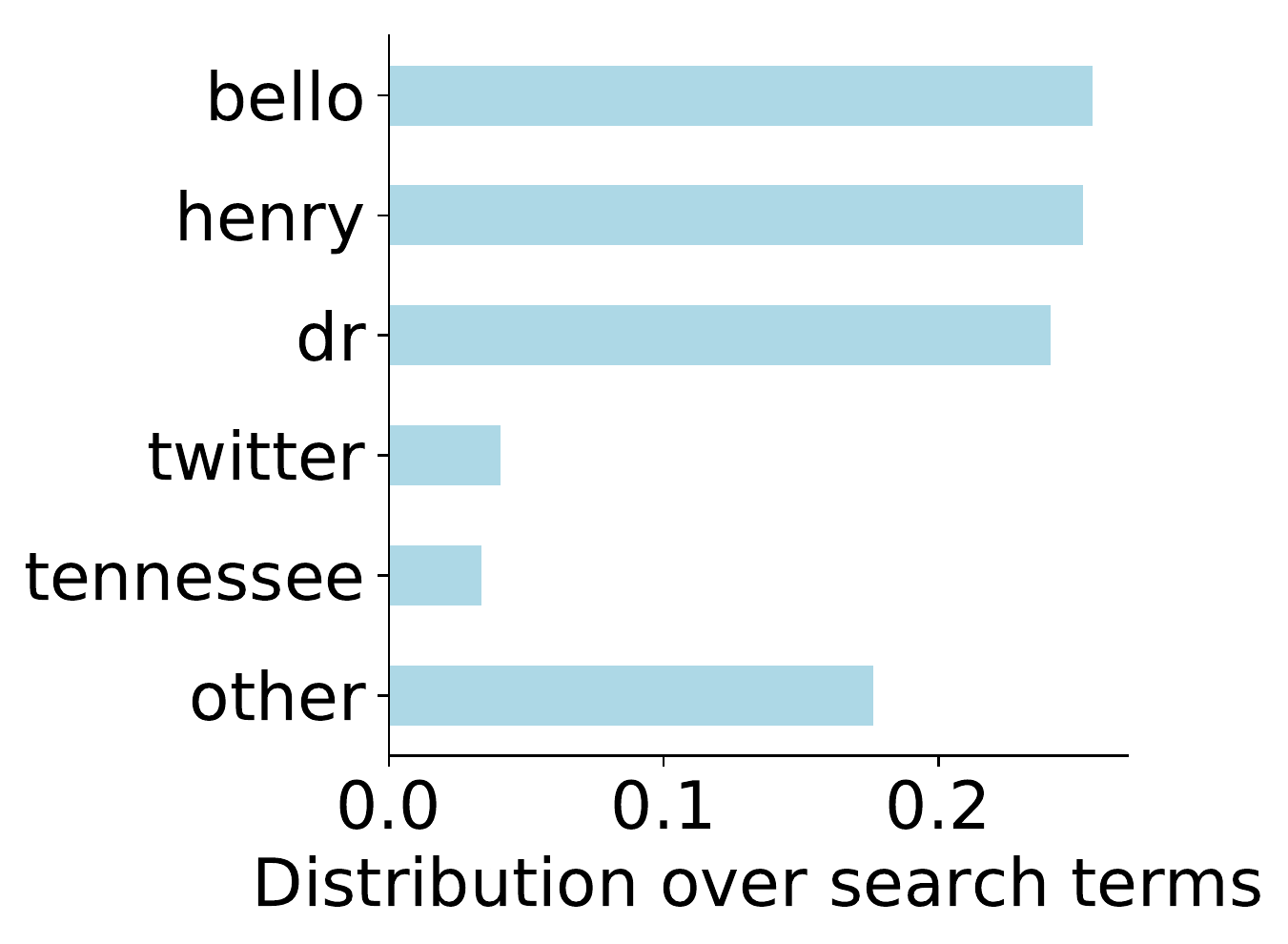} 
    }
\caption{\textbf{Case study: Tweet impressions for the story of Henry Bello.} Subfigure (a) shows impressions on IRA tweets. The surge in impressions is likely linked to the early availability of news information that was unavailable elsewhere; most news outlets only published stories on this subject the following day (Subfigure (b)).
Subfigure (c) shows the most common search terms that generated the impressions. Most of the stories were published by local news outlets.}
\label{Fig::HenryBellos}
\end{figure}

A news event about Dr. Henry Bello occurred on June 30, 2017. This turned out to be a major news story in New York City, receiving thousands of news articles of coverage \cite{gdelt}. However, the bulk of these articles were published and indexed by the search engine on July 1, 2017. Fifteen of the stories published on July 1st were published by the Associated Press, however the remainder came from local domains, with the top domains being: nydailynews.com (12 stories), denik.cz (12 stories),  heraldsun.com.au (11 stories) and
news-sentinel.com (10 stories). Because the IRA tweeted about the event around noon, they got indexed on June 30, beating many of the articles published and receiving a surge of impressions (i.e. be displayed as a search result) for the story (see Figure Figure ~\ref{Fig::HenryBellos}).

\subsection{Case study: \url{blackmattersus.com}}

We showed earlier that a large volume of apolitical content was produced by the IRA. We examine the \url{blackmattersus.com} domain as a case study of how this content content played a role in a larger strategic aim: search engine traffic followed by an on-site engagement funnel.

\begin{figure}[ht]
    \centering
    \footnotesize
    \begin{tabular}{cc}
       \subfloat[Emotion tags given by crowdsourced workers to \protect\url{blackmattersus.com} URLs, weighed by traffic from either Facebook or search.]{ \includegraphics[width=.18\textwidth]{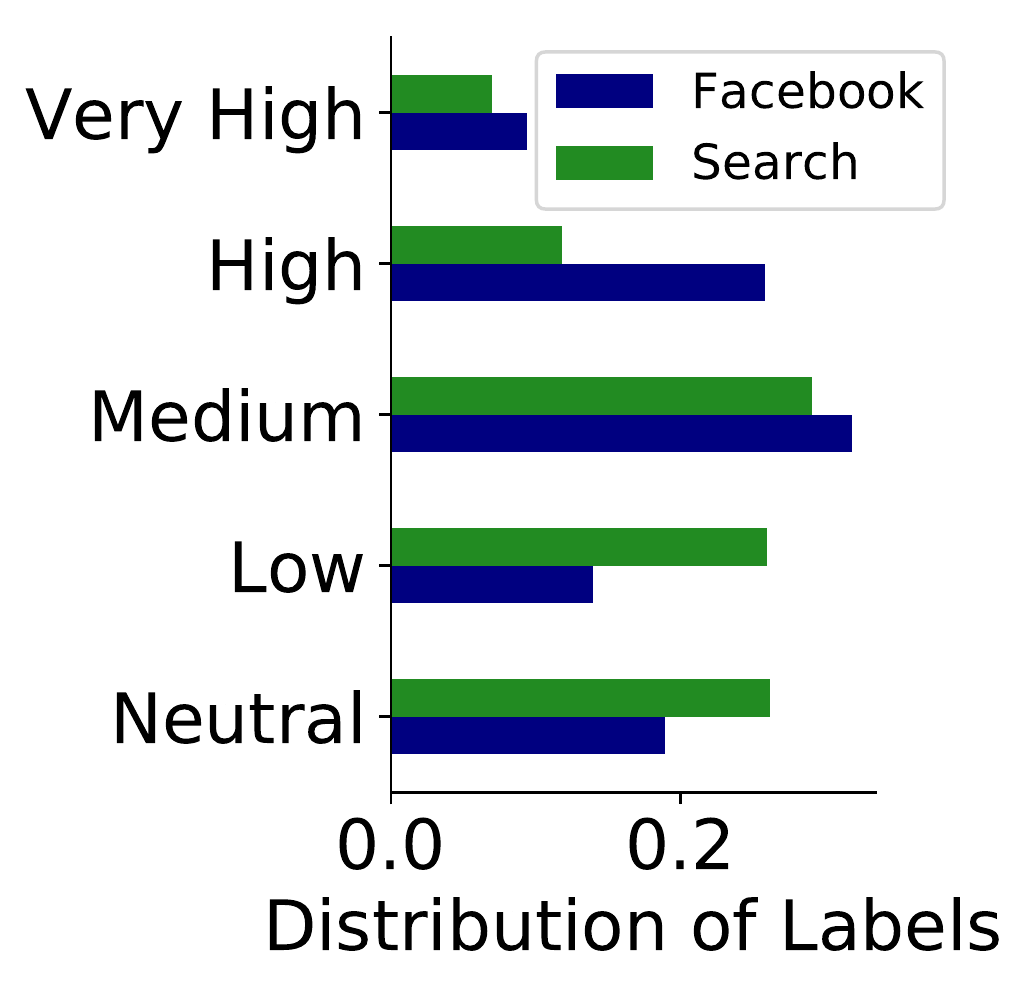}}
       & 
       \subfloat[Emotion tags given by crowdsourced workers to \protect\url{blackmattersus.com} URLs, weighed by traffic from either Facebook or search.]{ \includegraphics[width=.24\textwidth]{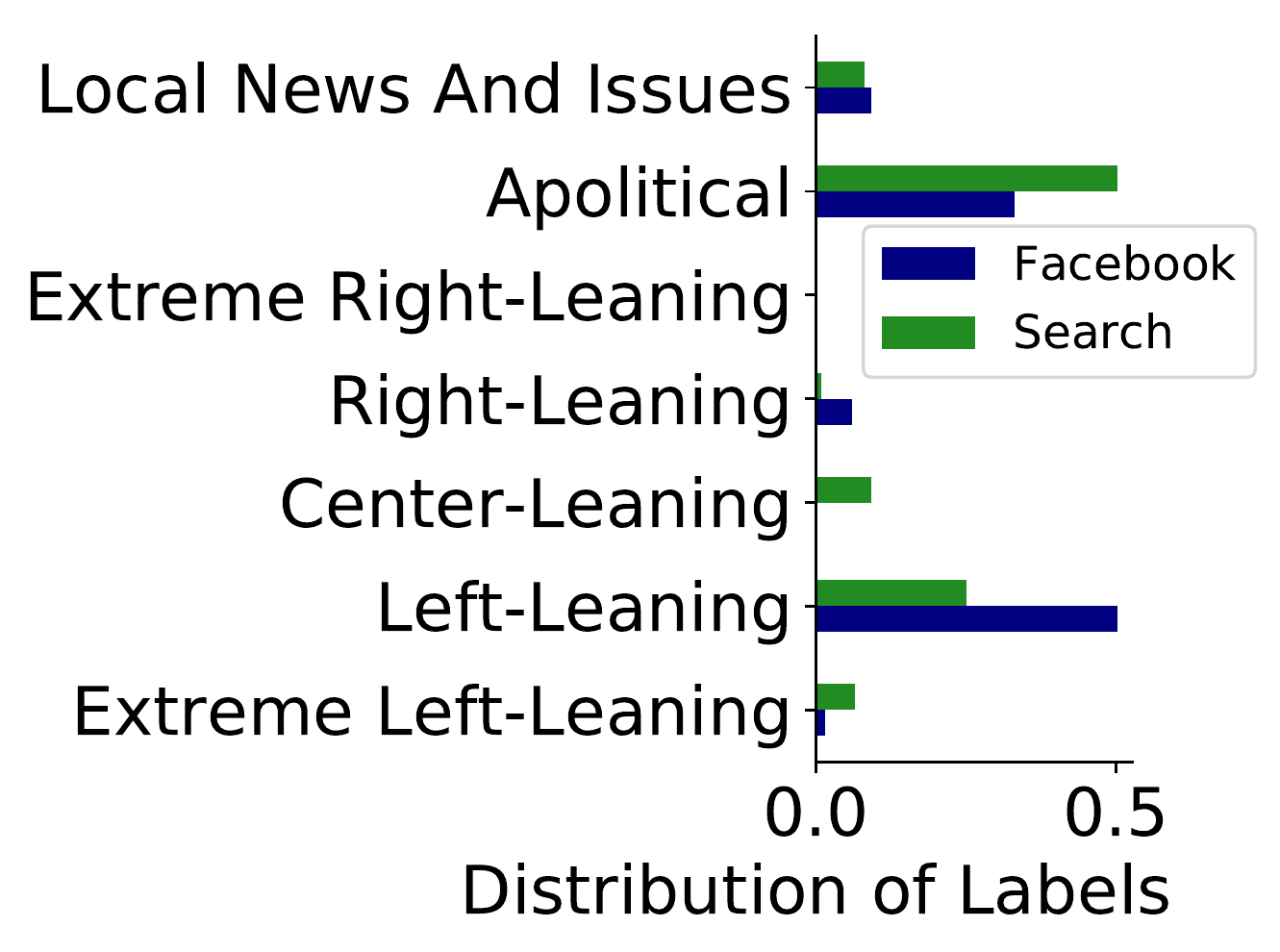}
       }
   \end{tabular}
    \caption{\textbf{Case study: Blackmattersus.com, summary.} \url{Blackmattersus.com} was the most highly trafficked IRA-owned domain in our dataset. The emotional intensity of pages users reached from Facebook.com is higher than that for pages reached via search, as can be seen by looking at the top-trafficked pages on the domain, as well all trafficked pages as labeled by crowd workers.}
    \label{fig:SearchSocialEmotions}
\end{figure}

\subsubsection{Search Traffic}

\url{blackmattersus.com} contained a mix of content: content that was apolitical and emotionally neutral, as well as content that was political and emotionally intense (Table ~\ref{table:blackmattersusexamples} and Figure ~\ref{fig:SearchSocialEmotions}). We found that content rated as ``Apolitical'' drew more traffic on Bing while ``Left-leaning'' drew more on Facebook. Additionally traffic to \url{blackmattersus.com} from Facebook was drawn more to pages that were ``High" or ``Medium" emotion-level, while traffic from Search centered on pages that were more likely to be rated ``Neutral" or ``Low" emotion.\footnote{These findings provide support to the speculation voiced by journalists at ~\cite{npr}, that the IRA may have been generating apolitical content to serve a purpose: to gain Americans' trust in a network of seemingly local news handles, which could later be ``operationalize[d] [to] significantly influence the narrative on a breaking news story".} We next explore the mechanisms \textit{behind} this content performing well through search.

\begin{figure*}[!htbp] 
  \includegraphics[width=\textwidth,height=6cm]{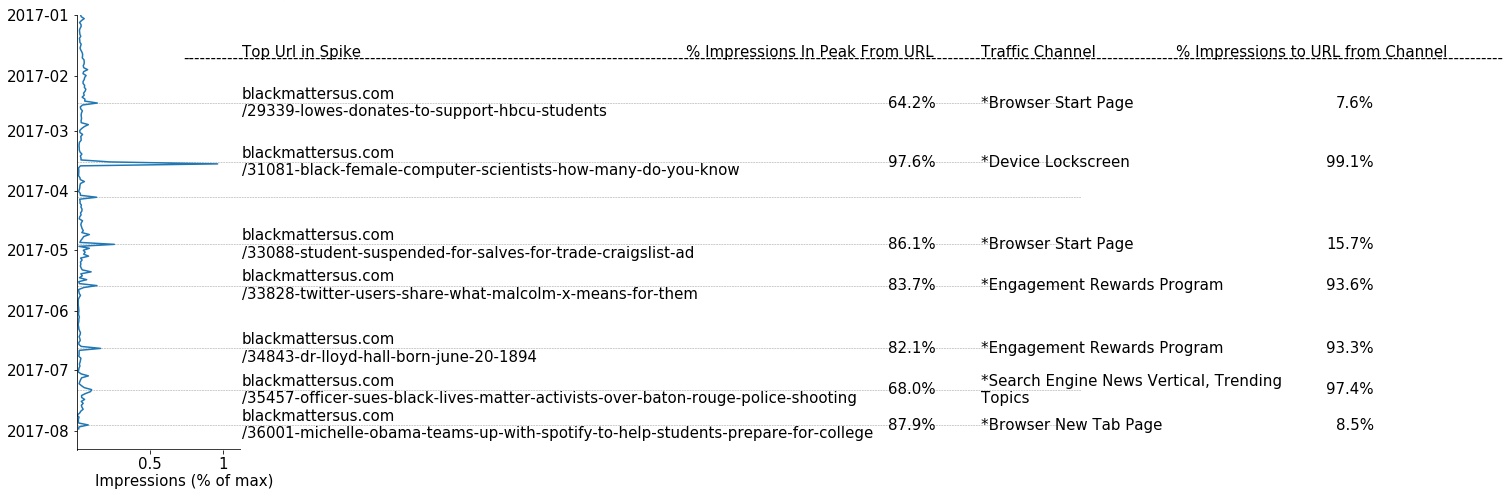}
  \caption{\textbf{Instances of Facebook-Promoted Domains Appearing in Search Results}: The domains most associated with the Facebook-ad campaign often appeared as results to search queries, with surges of search query impressions occurring from time to time. Some of these spikes can be associated with the promotion of various topics. For instance, the largest spike corresponds to the page  \url{blackmattersus.com/31081-black-female-computer-scientists-how-many-do-you-know} appearing among the results returned for the query, ``women scientists'' -- a query which was briefly promoted on the lock screen of some Microsoft devices.}
  \label{Fig::BingImpressionChannels}
\end{figure*}

Beyond generating ranked lists of results in response to search queries input by end users, major search engines also power a host of company properties like news verticals, homepages and new-tab pages, which can link out to preformulated queries like, for example, ``women scientists". Fig.~\ref{Fig::BingImpressionChannels} shows that, on several occasions, \url{blackmattersus.com} appeared in the results returned for such preformulated queries, which were promoted across several Microsoft properties, including \url{bing.com/news} news verticals, new tab pages, and device lock screens.

By publishing neutral, entertaining content 
the IRA was able to draw more users to its pages.

\begin{table}[]
    \centering
        \begin{tabular}{|l|}
            \hline
           \textbf{ Top Clicked URLs via \url{facebook.com}} \\
            \hline
            \hline
            /nj-police-officer-arrested-and-charged-f... \\
            /her-provocative-ballet-dance-hurts-white... \\
            /tamron-hall-has-been-replaced-by-megyn-k... \\
            \hline
            \hline
            \textbf{Top Clicked URLs via Search} \\
            \hline
            \hline
            /black-history-month-black-inventor... \\
            /katherine-johnson-a-black-nasa-pioneer... \\
            /black-female-computer-scientists... \\
            \hline
        \end{tabular}
    \caption{\textbf{Case study: Blackmattersus.com, examples.} Top three most trafficked pages on \url{blackmattersus.com} from either Facebook (top) or Search (bottom).}
    \label{table:blackmattersusexamples}
\end{table}

\begin{table}
\centering
\footnotesize
\begin{tabular}{|l|r|r|r|r|}
\hline
{} & Share Button& Contact Page & Meetups & Donate \\
\hline
\hline
\# links & 3940 & 1 & 896 & 1 \\
\hline
Click rate & 2.64\% & .28\% & .11\% & .056\% \\
\hline
\end{tabular}
\caption{\textbf{Blackmattersus.com Structure:} 
``Share Button'' are links on each article that encourage sharing on social media, the ``Contact Page'' allows users to sign up for email correspondence, ``Meetups'' is a listing of local meetups and ``Donate'' is a page soliciting donations that leads to a \url{paypal.com} page. (We saw no evidence of clicks to \url{paypal.com} in the same session.)}
\label{table:blmstructure}
\end{table}

\subsubsection{Engagement Funnel}

While gaining a wide reach through various promotional channels on the internet, \url{blackmatters.com} was structured to potentially engage users beyond reading articles.  

\url{blackmattersus.com} is a particularly expansive site with $4,608$ pages: articles, pages linking to Facebook event pages, a donations page, and an input field soliciting users to join the site's mailing list. Shortly upon arrival, a pop-up asks users if they would like to receive notifications. 

We performed a site-wide scrape to categorize pages hosted on the domain. 
Table ~\ref{table:blmstructure} gives our overview of this domain. On the first row, we show the raw number of links we counted devoted to each of four action types: ``Share Button'', ``Meetups'', ``Contact Page'', and ``Donate''. Interestingly, the ``Meetups'' are pages onsite that link out to \url{meetup.com} and Facebook events, many of which appear to be legitimate events, and are still active on Facebook. The ``Donate'' button links to a \url{paypal.com} page\footnote{The \url{paypal.com} page is operated by a merchant with the email address \textit{xtimwaltersx@gmail.com}, which was one of the emails mentioned in the Special Counsel's Indictment \cite{mueller_ira_indictment}.}.

Each of these link-types is accessible from any article. The number of users we see engaging with this funnel is very low. The second row of Table \ref{table:blmstructure} shows the conversion rate, or the percentage of users who clicked on one of the links \textit{after} clicking on another blackmattersus.com page. These numbers are only one 1/10th typical conversion rates observed across e-commerce sites industry-wide \cite{conversion_rates}.

\subsection{Case study: Did promotions matter?}

\begin{figure}[!ht]
\centering
\begin{tabular}{l|l}
    \url{stopallinvaders}: & 
    \url{brownunitedfront}: \\ 
    \hline \\
    \subfloat[Spikes in unique users per day (black) sometimes co-occur with photo and video uploads throughout the period (green).]{\includegraphics[width=.22\textwidth]{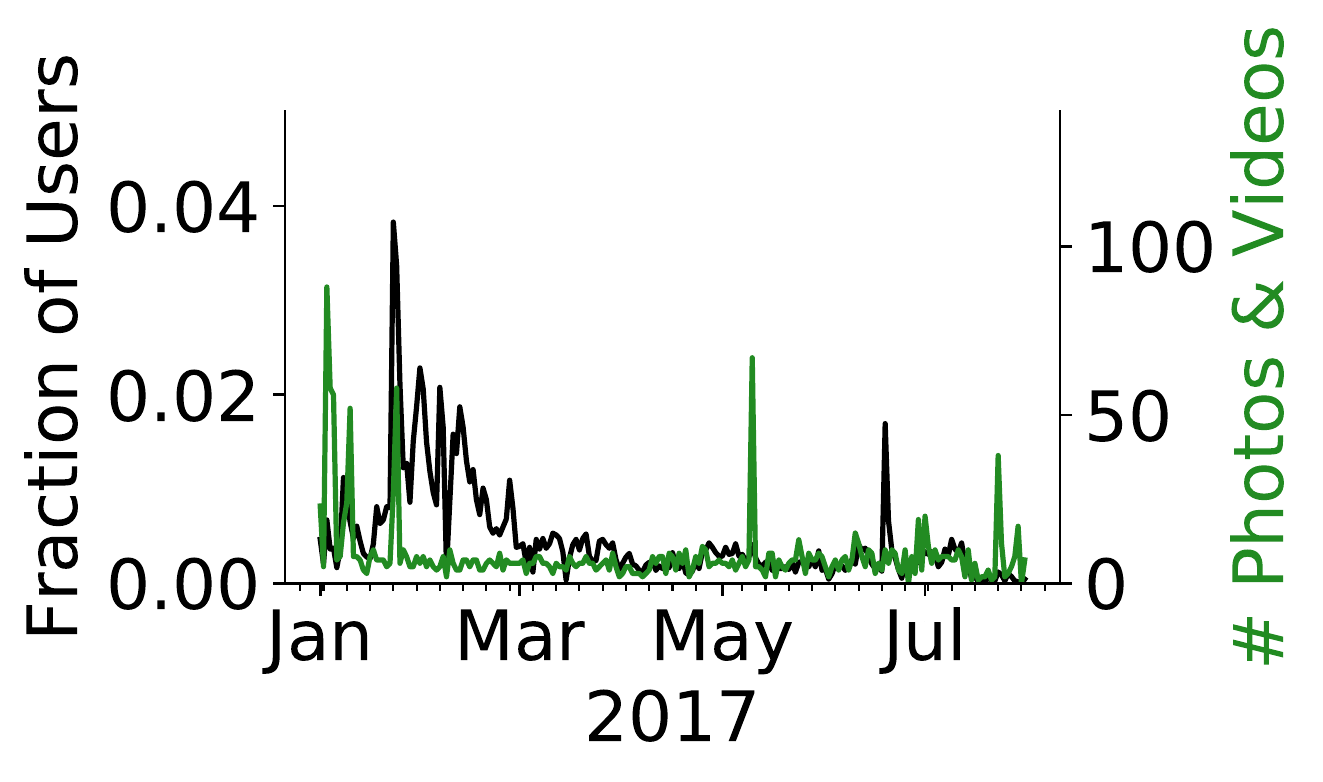}} &
    \subfloat[Spikes over time on this property, in contrast, do not show any clear relation to photo and video posts.]{
    \includegraphics[width=.22\textwidth]{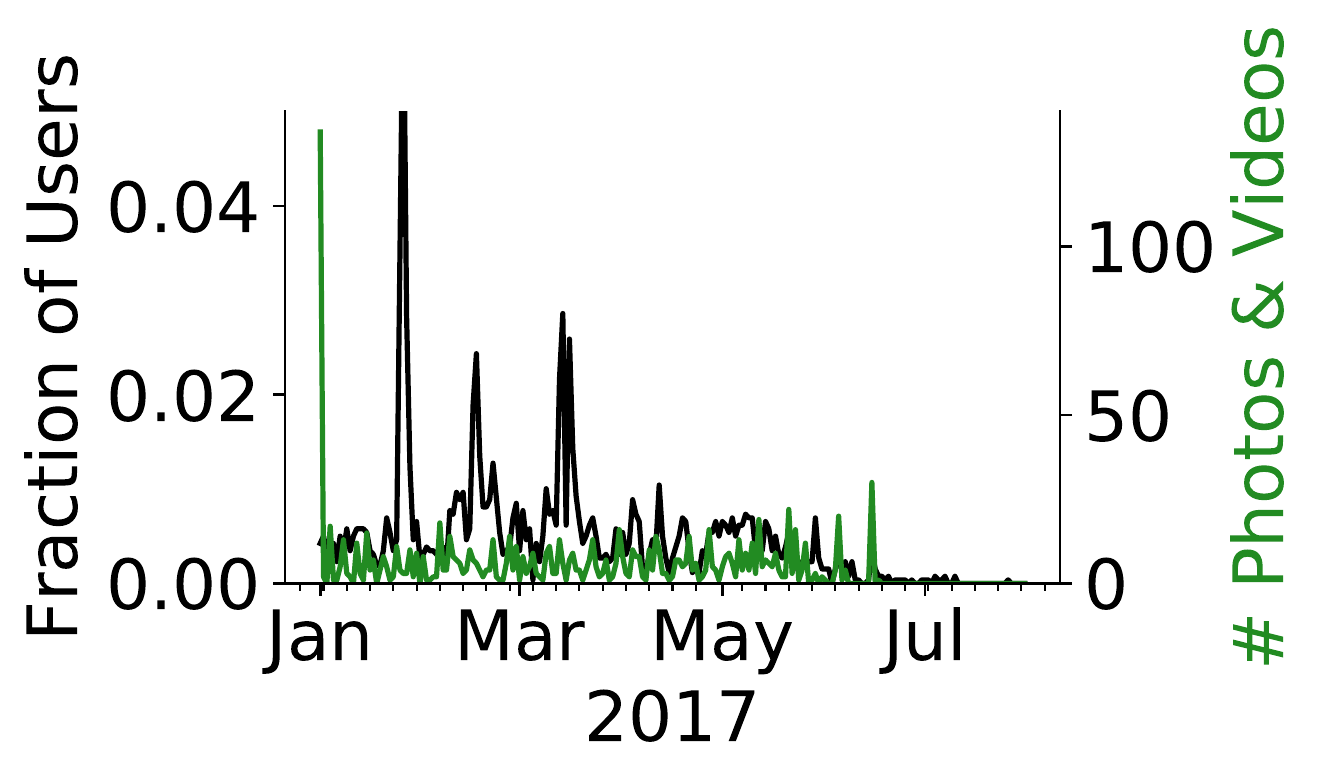}} \\
    \subfloat[Spikes in unique users per day (black) show no alignment with promotions (maroon).]{
    \includegraphics[width=.22\textwidth]{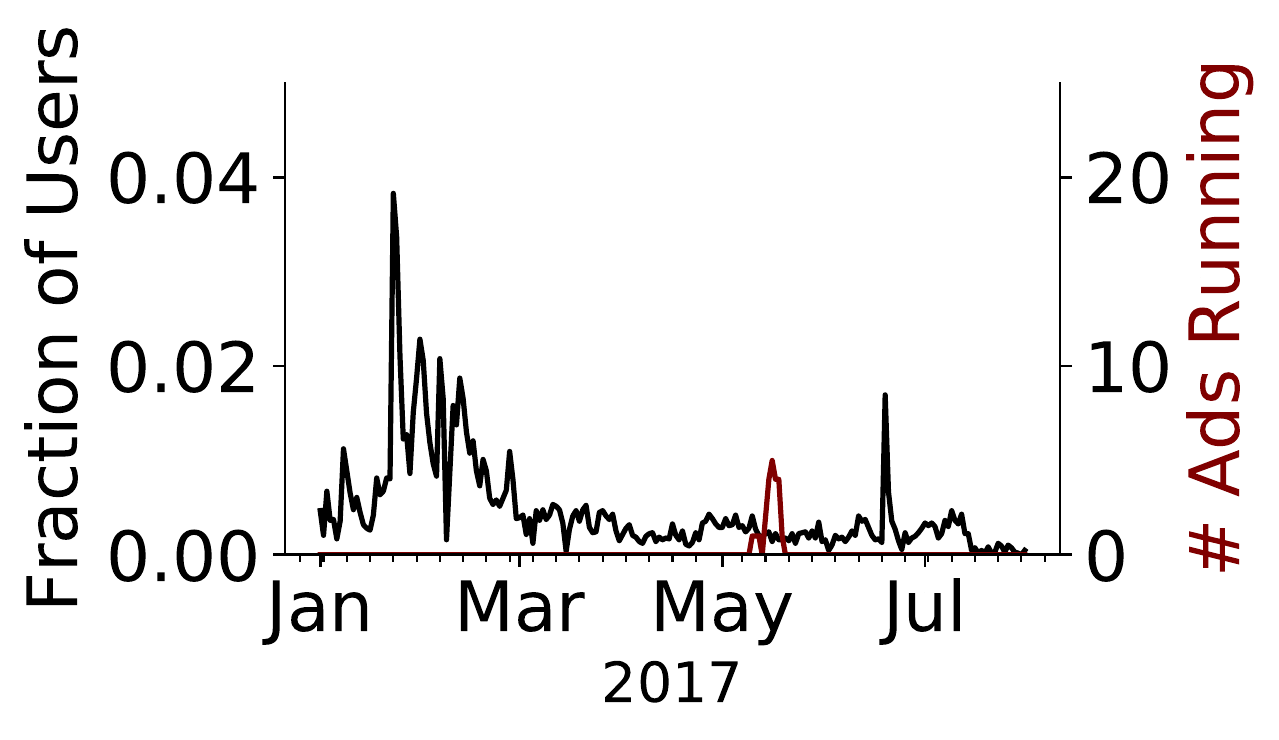}} 
    &
    \subfloat[Spikes in unique users here show alignment with promotions.]{
    \includegraphics[width=.22\textwidth]{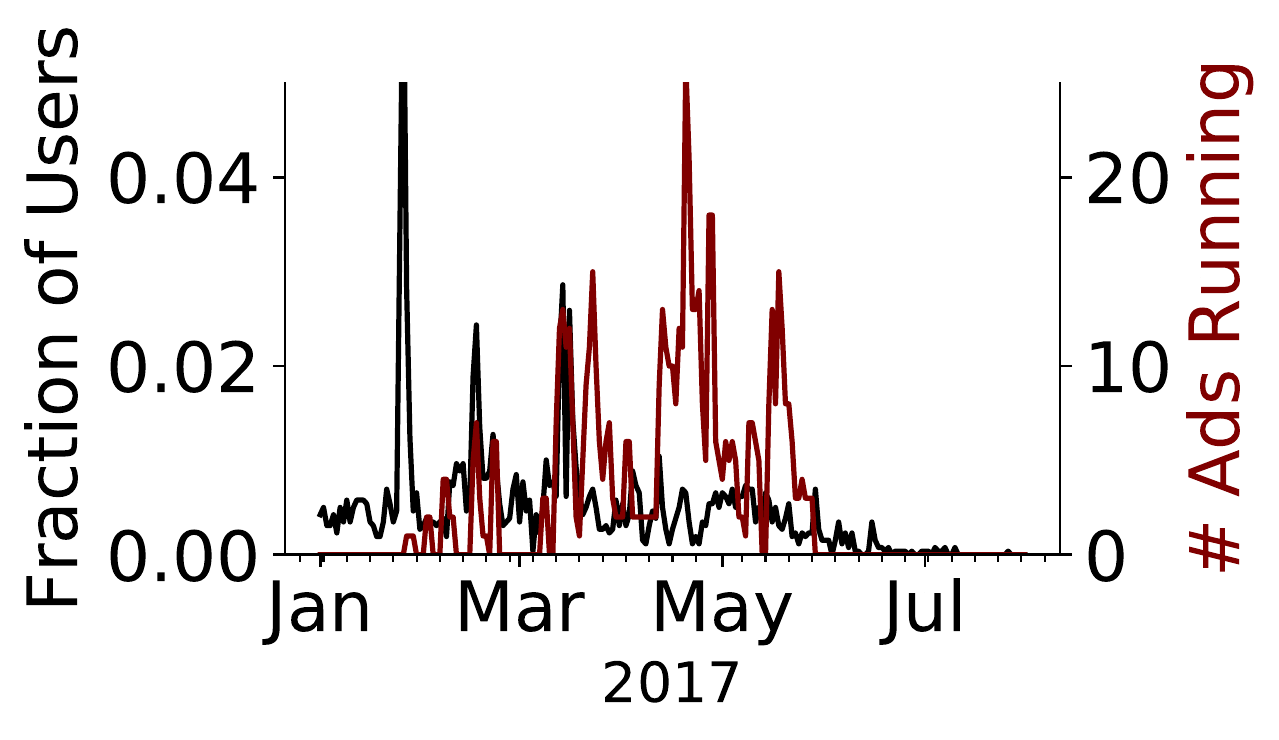}
    } 
    \\
\end{tabular}
    \caption{\textbf{Drivers of Spikes.} Many promoted web-properties (shown above) seem to have click-spikes associated with factors like paid actions (Facebook-ads) or unpaid actions (photo or video uploads). The Y-axis in each of the graphs represents the fraction of total unique users to the group over the time period that visited on a given day. We quantify the effect of a paid Facebook promotion on the day of promotion by predicting spikes the property would have received had it not been promoted (see Appendix 2). This figure shows that in addition to paid promotions, unpaid actions (such as photo/video uploads) also impacted the traffic to the Facebook properties.}
    \label{Fig::PromoEffects}
\end{figure}

The IRA's ad-promotion strategy on Facebook involved roughly 531 ads and nearly $1$ million rubles (or $\$15,000.00$) on left-leaning content, compared with 184 ads and $620$ thousand rubles (or $\$9,112.00$) on right-leaning content.
The IRA spent more on left-leaning content during our observation period.

Clearly, paid advertisements were part of the IRA strategy to generate traffic to their groups and properties. However, paid promotions are only part of the story. There are other important known and unknown factors not to be ignored. Among the known ones: historical traffic, photo and video shares are the most correlated ones with traffic spikes.

For example, Fig.~\ref{Fig::PromoEffects}(a)-(b) illustrates the traffic to different Facebook groups over time along with (1) photo and video uploads and (2) paid promotions. For the StopAllInvaders Facebook group, it seems some traffic spikes co-occur with events like photo and video uploads, but do not co-occur with paid promotions. For another example group, BrownUnitedFront, spikes in traffic seem more aligned with paid promotions than with media uploads. To isolate the influence of paid promotions from other unpaid factors, we trained a model with the goal of predicting the changes in spikes of high-traffic Facebook groups in the absence of paid promotions.

We used our model to examine six Facebook groups chosen for having high traffic as well as a large number of promotions. For these groups, we infer that without paid promotion, the left-leaning groups would have received fewer traffic spikes, while traffic spikes for right-leaning groups would have been largely unchanged. However, an important take away here is that paid promotions were not the only factor that lead to traffic spikes, which is something to keep in mind for future work.

A more comprehensive discussion of this is included in Appendix 2. 
 \subsection{Geographic patterns in traffic}
87\% of the traffic to Facebook-promoted properties was U.S. based, but we observed clicks to IRA's properties from 176 countries. 

We consider a broad array of geographic features to describe the areas contributing traffic including race-based, rural/urban, education-based, employment-based and other demographic features from the 2010 census \cite{census2010}. We look at voting patterns from the 2016 presidential election \cite{votes2016}. We also examine voting as a percentage of registered voters, a metric that we use to proxy voter enthusiasm \cite{voterregistration2016}.

\begin{figure}[!ht]
\centering
    \subfloat[Traffic to Blacktivists Facebook group normalized by average daily active users by state.]{
    \includegraphics[width=.21\textwidth]{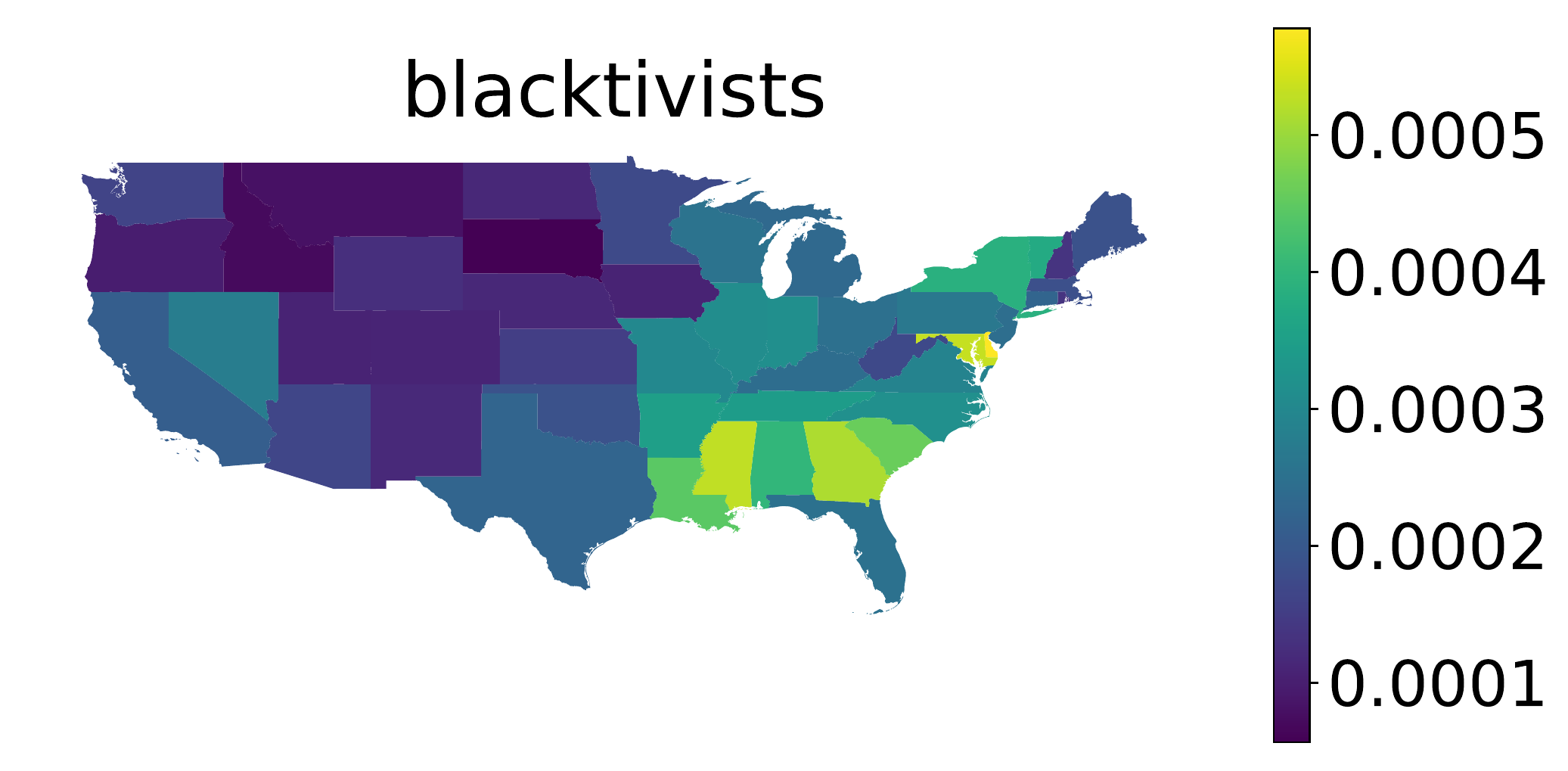} 
    } \hspace{1mm}
    \subfloat[Traffic to Secured Borders Facebook group normalized average daily active users by state.]{
    \includegraphics[width=.21\textwidth]{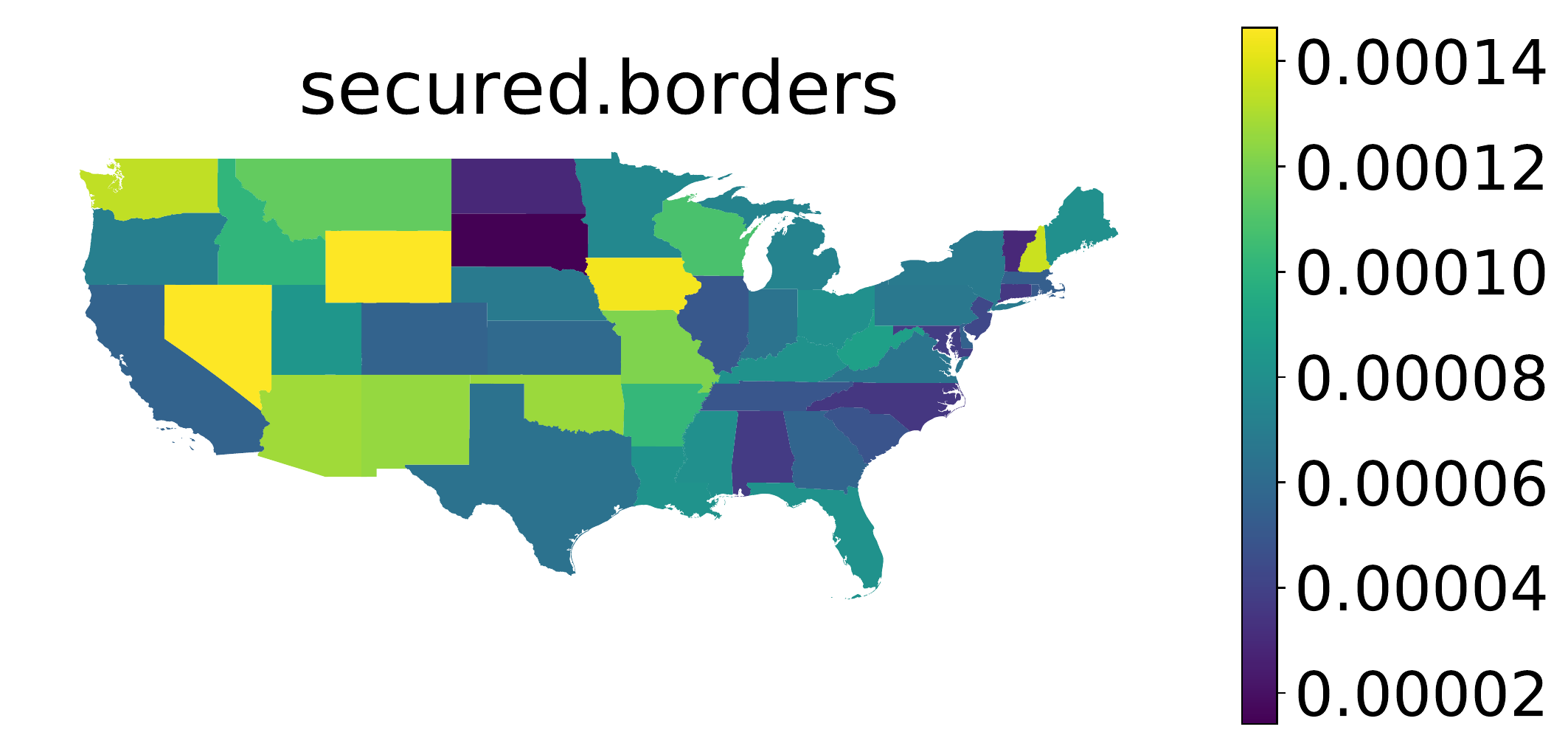}
    }\\
    \subfloat[Correlation coefficients across demographic census features with categories of the Facebook advertisements  ($p<0.1$ shown)]{\includegraphics[width=.45\textwidth]{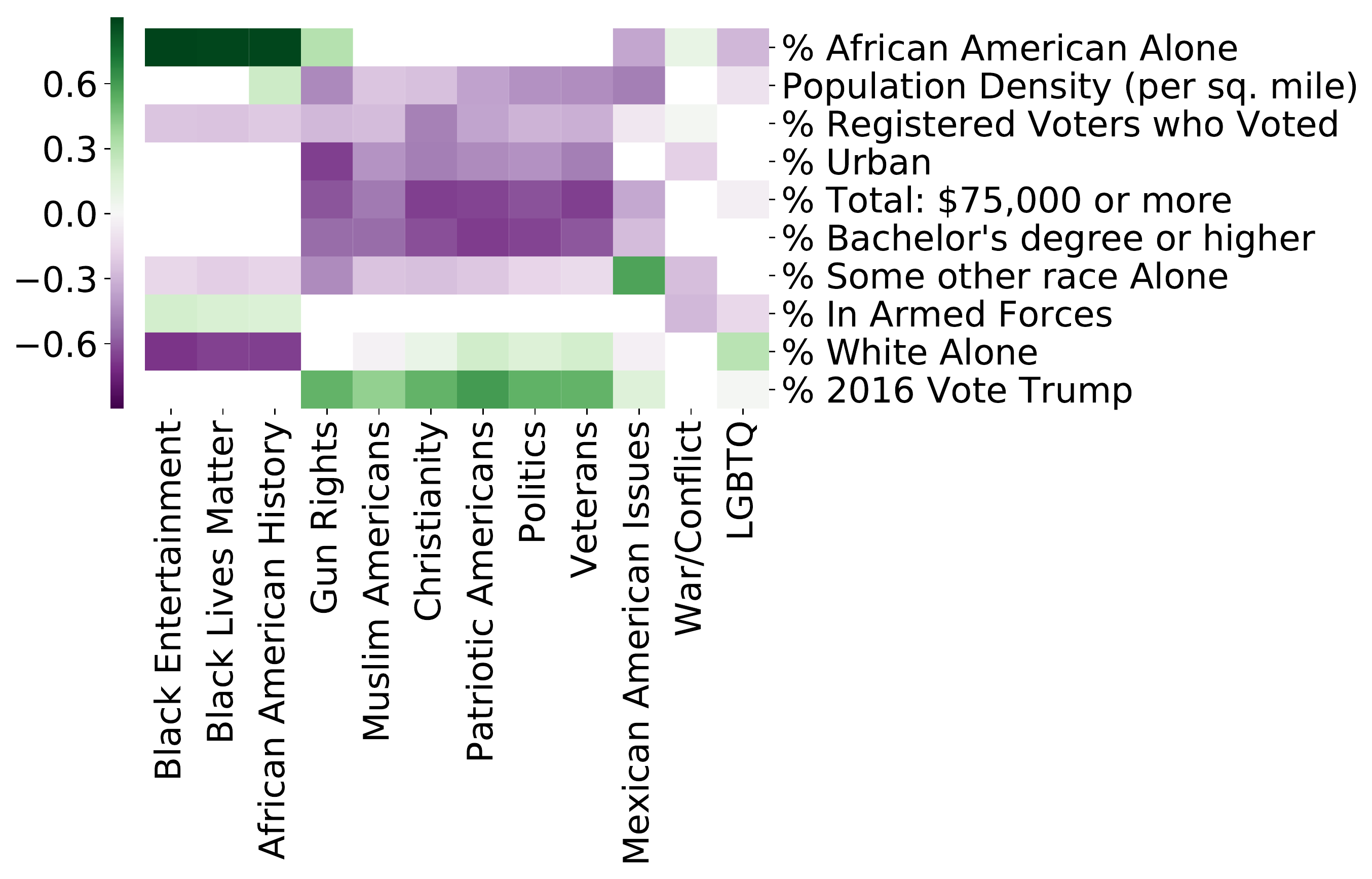}
    }
    \caption{\textbf{Correlation with state-level census features.} The Facebook-ad campaign was targeted primarily through interest-tags. We grouped them into clusters by asking crowd workers to label ad content. The top row shows distributions over traffic-normalized click counts by state for two sample groups. The bottom figure shows correlations between census features and normalized clicks by state ($p<0.1$ shown).}
    \label{Fig::StateLevelCorr}
\end{figure}

Fig.~\ref{Fig::StateLevelCorr} shows state-level correlations between census features and traffic to IRA properties promoted via Facebook. The IRA properties are clustered by targeting keywords and topics as per the results of the crowdsourcing study. We only show a subset of features, and only display correlations with $p<.1$.

Traffic to IRA content in certain categories is positively correlated with demographic features. For example, traffic to IRA content labeled ``Black Lives Matter'' is significantly correlated with the percentage of census respondents self-identifying as African American in a state ($c=.88$, $p<.001$). On the other hand, this content is negatively correlated with the percentage of census respondents self-identifying as White in a state.
 
Moreover, there is a significant negative correlation ($c=-.29, p<.001$) between the percentage of of Registered Voters who Voted (in 2016) in a state with traffic across most ad groups. Thus, there was more traffic to ads in states with lower voter turnout. This may support the claim that the IRA operation was seeking to target areas of voter discontent~\cite{shane_nytimes}, but is subject to many confounders. Furthermore, it is important to note that our work is looking at data collected in 2017, two months after the 2016 U.S. Presidential election, and thus cannot infer anything about the pre-election timeframe.
 \section{VI. Effects on exposed users}

Having described case studies of the IRA strategy, 
we now close with a view of the effects on users. Although our main conclusion here is that few effects were observed, more research is needed due to the following limitations. First, this analysis cannot observe the long-term behavior of internet users online due to extensive anonymization techniques applied in the data. Second, browser and query logs give us only a limited view into users' online actions (e.g. we cannot see posts, tweets and emails), and most importantly we cannot observe real-world behaviors. Third, a complete analysis on these effects calls for data from potentially multiple browsers and search engines to better cover the online user base, as well as data from a broader range of mobile devices and applications. For completeness, we summarize all attempts we made to better understand impact on exposed users and hope that these attempts will seed further investigation on the topic.

Our study explored online news consumption, online political donations, and real-world protests. 
To categorize online news consumption, we used data from MediabiasFactcheck.com to label domains known for publishing various news-types ~\cite{mediabias}. Using these labels, we examined the browsing behaviors of users exposed to an IRA ads before versus after exposure to the ad, while controlling for session length. We did not see significant changes in the total volume of news consumed, or in the extremism of news sites visited before and after exposure to the ad.

To understand online political donations, we catalogued the websites registered for candidates, politicians and major recipients of political donations, using data from OpenSecrets.org \cite{opensecrets} and BallotPedia.org \cite{ballotpedia}. Again, we failed to see any significant changes in pages visits before and after exposure to an IRA ad. 

In addition, we also explored the notion that the IRA was attempting to draw more people into protests. For example, Robert S. Mueller's indictment points to anecdotes of Americans being driven to attend real-world protests as a result of misinformation \cite{mueller_ira_indictment}. Using Crowd Counting Consortium's numbers tracking protests in 2017, we examined whether there was a correlation between where protests occurred in the U.S. and the locations of ad-clicks by users \cite{crowdcounting}. After normalizing for state-level populations and state-level browsing behavior, we found no significant correlation.
 
\section{VII. Related Work}
\noindent {\bfseries Social media and politics.} 
The dataset generated by this paper was further used in followup work by Boyd, Spangher et al. \cite{boyd}. This work examined language differences in the IRA tweets and those posted by a general U.S. population during the same time. Additionally, Boyd et al., report that the majority of Facebook ads were posted during 9am and 6pm, Moscow Standard Time. As such, \cite{boyd} provides further analysis for the point-of-origin of IRA tweets and advertisements. 

The impact of social media on the political ambit has attracted significant attention more broadly in both political science and social media analysis studies. 
Social networks have facilitated faster and targeted user communication, which has created new and amplified effects due to high information diffusion \cite{bakshy2012role}. From a political perspective, such effects can often be positive and increase political participation awareness \cite{gil2012social}.  
However, as any powerful means of communication, social platforms can also be exploited by external entities to produce negative effects in the society such as drastic political division~\cite{sunstein2018republic} as well as manipulation and misinformation~\cite{marwick2017media,fourney2017geographic}. This paper specifically focuses on extracting data-analytic insights that can reveal information regarding the extreme and potentially divisive characteristics of the IRA campaign. Being aware of the nature and the scale of these characteristics is crucial for increasing general awareness and developing robust protective mechanisms in the future.

\vspace{0.33em}
\noindent {\bfseries Division and polarization.}  Political division has been investigated using the notion of ``information bubbles" \cite{resnick2013bursting} and ``echo chambers" \cite{DBLP:journals/chb/Bessi16,DBLP:conf/www/GarimellaMGM18}.
Authors study the phenomenon of political and intellectual isolation that can occur when users are only exposed to information that confirms their own beliefs.  While some of these effects can exist due to organic user behavior, recent research has shown that they have been leveraged to increase polarization and division \cite{stewart2018examining,conover2011political,DBLP:conf/icwsm/GarimellaW17,DBLP:journals/chb/Bessi16}.
In this context, various tools have been proposed to foster diverse information consumption either via visualization ~\cite{DBLP:conf/www/GillaniYSVR18} or exploration strategies that go beyond the personal network bubble~\cite{resnick2013bursting}. 

\vspace{0.33em}
\noindent {\bfseries Manipulation and misinformation.} Prior research---including our own---has found social media can be used to to spread and amplify propaganda at a global scale \cite{menczer2016spread,fourney2017geographic,silverman}. Combined with the fact that humans are inherently bad at detecting deception~\cite{mihalcea_automatic_2013,abouelenien_deception_2014,perez-rosas_deception_2015}, 
decades of algorithmic advances in targeted advertisement have laid the groundwork for large-scale, general-purpose mass manipulation.

Manipulation and misinformation strategies transform information in a form that best propagates ideas or agendas of specific groups of interest. A comprehensive summary of these techniques~\cite{marwick2017media} shows several examples where social media has shown to be vulnerable to such attacks especially from extremist political groups and Internet trolling entities. In the most harmful form, manipulation can be encountered as jointly combined with misinformation, where information is intentionally twisted or even fabricated. Social media has been a lucrative target of misinformation~\cite{allcott2017social,fourney2017geographic,faris_partisanship_2017} in the recent years. 
Social bots on Twitter have played a role in influencing the spread of information in a network~\cite{varol_online_2017,varol_early_2017,bessi_social_2016}.
Various approaches have been proposed to isolate and limit the harmful effects of misinformation  by harnessing linguistic features, semantic analysis of the content, and network topology properties ~\cite{shu2017fake,budak2011limiting,conroy2015automatic}. 

\vspace{0.33em}
\noindent {\bfseries The 2016 US election.} The events of the IRA ad campaigns in the US presidential election in 2016 are not unique. Similar developments have been noticed in the last five years in the case of the Brexit referendum in the UK in 2016~\cite{howard2016bots}, German Federal election in 2017~\cite{morstatter2018alt}, and elections in Pakistan in 2013~\cite{younus2014election}. This work was motivated by the need to better understand the IRA campaign in 2016 and was enabled by the release of the recent datasets from Facebook and Twitter on this matter during the respective testimonies at the United States House of Representatives~\cite{schiff_facebook,schiff_twitter}. Guided by the same 
motivation, there is recent and ongoing work tackling important questions related to the structure of the retweet network~\cite{stewart2018examining} and the characteristics of IRA-promoted Twitter handles~\cite{clemson}. 
Multiple journalistic efforts~\cite{nytimesfb,npr} published illustrative examples of the promoted content and informed the broader audience. This paper aims at providing a data-oriented analysis, joining multiple sources of information funneling from the promoted ads and content, to post-election web traffic, to observable user effects.
 \section{VIII. Discussion and Conclusion}
While our studies present a set of insights on the strategy, structure, and scope of IRA-related web activities, our analyses have limitations. We highlight these as caveats and as guides for future work in pursuit of confirmation of the results and inferences shared in this paper. First, much of our analysis relies on browsing instrumentation data on desktop computers and excludes mobile devices. The released Facebook ads data set and the web search data set we use, on the other hand, do include mobile traffic. 
Secondly, because the browsing instrumentation collects only URLs, our browser logs cannot directly observe interactions that occur within a page (e.g., enabled by AJAX or JavaScript). We inspected many of the web properties and, while we found that most activities of interest result in changes to the URL, some (e.g., donations brokered by Facebook) could not be detected. 
Also, as with all studies of this nature, we cannot observe actions taken in the physical world. For example, we cannot be sure if a person attended a protest that they read about. Likewise, we restrict our analysis to short-term trends because our browsing instrumentation data set does not maintain long-term histories. Our data is also collected in 2017, after the 2016 election, which was one of the targets of the IRA campaign. 

Nevertheless, we have demonstrated multiple ways by which malicious agents can manipulate the digital landscape showcasing the large attack surface susceptible to malevolent interventions. We highlight how Facebook's fine-grained interest-based targeting keywords can yield emergent demographic-based information flow. We show
how search platforms, and their agnostic content-promotion strategies can also be leveraged -- especially when agents are able to get their late-breaking content indexed before more reputable sources.
We hope this work will further stimulate the development of new approaches to monitoring, understanding, and uncovering propaganda campaigns as well as technology and policy-based solutions that help avoid political manipulation. 
Our findings suggest that cross-organization collaboration will be valuable in this endeavor.

Circumstances around the 2016 U.S. presidential elections and rising concerns about the influence of propaganda campaigns led to the public availability of valuable datasets for understanding IRA activities. Weaving together the public datasets with proprietary data on search and browsing activity provides a previously unavailable lens on the workings of the IRA campaign. We consider this work a preface to numerous opportunities ahead and to the many directions that remain to be explored. We see a dual moving forward, with the Web jointly holding great promise for strengthening liberal democracies while also serving as a platform that can be harnessed by those who seek to manipulate and disrupt.  As the digital world continues evolve, and risks to democracy continue to emerge, openness and cooperation among major stakeholders will be essential to understand and counter malevolent threats.
 \bibliographystyle{aaai}
 {\small \bibliography{bibliography}}
 \clearpage
 \appendixpage
 \section{Appendix 1: Campaign details}
\subsection{Campaign Structure and Content}

In this appendix, we describe the campaign in more detail and discuss how our observations match with statistics released by Congress (\cite{schiff_facebook}, \cite{schiff_twitter}).

In the post accompanying the release of Facebook and Twitter data by the U.S. House of Representatives Permanent Select Committee on Intelligence\footnote{https://democrats-intelligence.house.gov/social-media-content/}, Representative Adam Schiff notes:

\begin{quote}
    ``The Russians ... weav[ed] together fake accounts, pages, and communities to push politicized content and videos, and to mobilize real Americans to sign online petitions and join rallies and protests.''
\end{quote}

We presented evidence of examples of the IRA promoting protests, meetups, and numerous engagement techniques with their ads, as discussed above. However, we did not find evidence of extensive user engagement. On the other hand, the content that we analyze in our paper contains some gaps and discrepancies with the content Schiff describes, which we explore and justify here. 

\noindent\textbf{Facebook} Rep. Schiff reports identifying the following content for the IRA's Facebook campaign \cite{schiff_facebook}:
\begin{itemize}
    \item $3,519$ Facebook ads.
    \item $470$ IRA-created Facebook pages.
    \item More than $80,000$ pieces of organic content posted to Facebook accounts.
\end{itemize}

We observe $3,519$ released PDF documents, each describing a Facebook ad, but as noted in Section II, we were only able to extract and process usable content from $3,061$ of these, i.e. $84\%$ of the posted ads as a result of blank ad-description fields, missing URL fields, blank documents, and OCR errors.

As noted in Section III, we only find $104$ IRA-linked Facebook groups being promoted by the ads, in contrast to \cite{schiff_facebook}'s stated $470$ groups. Formal communication and fact-checking with researchers at Facebook have confirmed that only $107$ of the groups identified were advertised. The missing $3$ groups in our data is likely due to OCR errors introduced when parsing released Facebook-ad PDFs. 

\begin{figure}[t]
    \centering
    \begin{tabular}{cc}
    \subfloat[][Total photos posted by account. A photo upload is determined by the minimum timestamp of a click on it's unique URL key.]{\includegraphics[width=.24\textwidth]{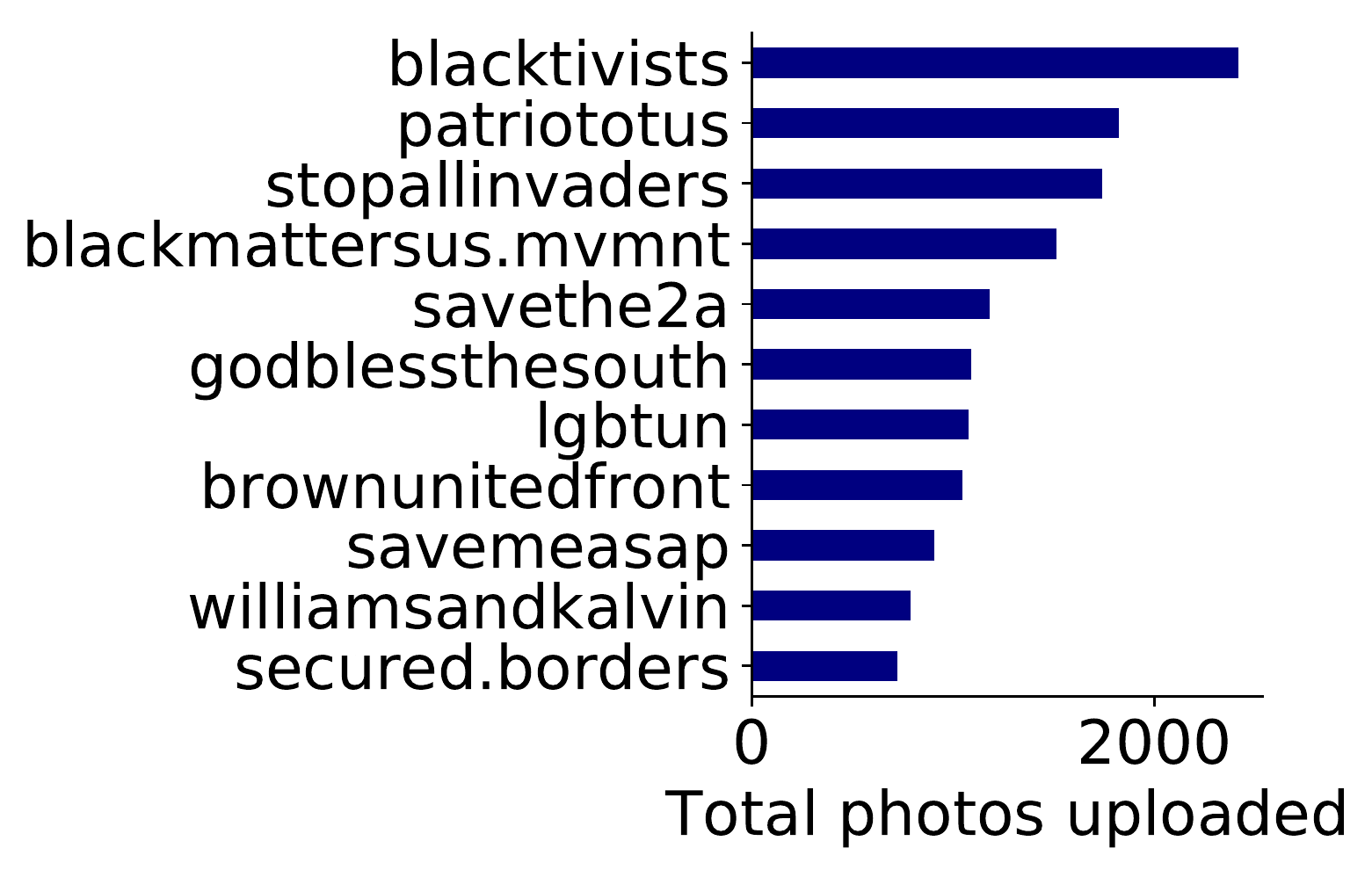}} &
    \subfloat[][Total videos posted by account. A unique video upload is determined the same way a photo upload is.]{\includegraphics[width=.24\textwidth]{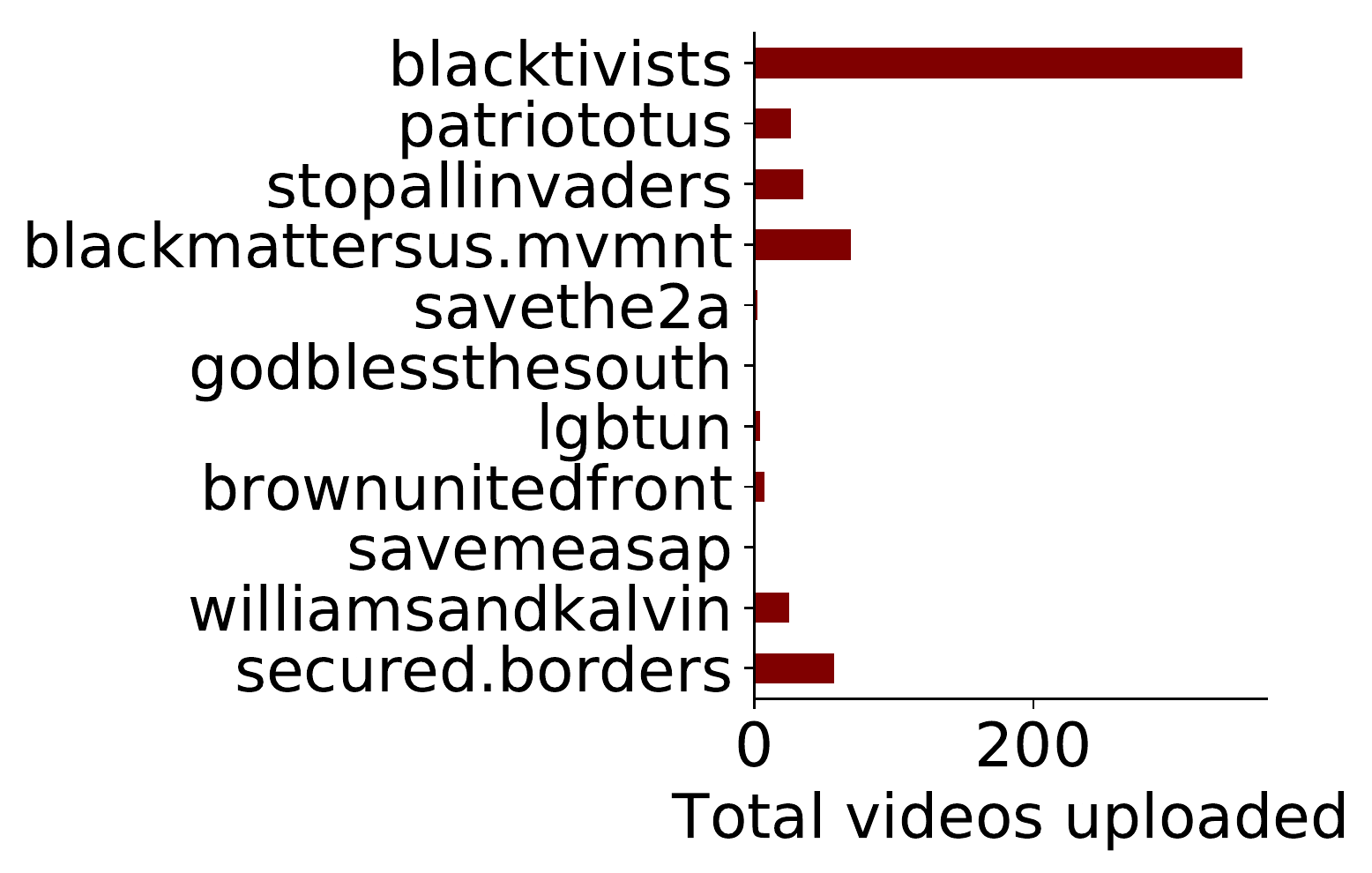}}
    \end{tabular}
    \caption{\textbf{Facebook photos and video uploads:} Metrics capturing the volume of organic (unpaid) posts involving photos and videos uploaded to IRA Facebook groups.}
    \label{fig:photosAndVideosCount}
\end{figure}

We seek to verify the volume of content reported by Rep. Schiff by deduplicating unique content URLs.  
We find a total of $14,860$ photo and video posts to IRA-linked accounts, shown in Figure \ref{fig:photosAndVideosCount}. We were not able to successfully track other post-types due to the nature of our instrumentation logs, and the manner by which other content appears on Facebook. It is likely that out of the $80,000$ pieces of content reported by \cite{schiff_facebook}, the over $65,000$ pieces that we did not observe included both: organic posts without photos and videos posted to the 104 IRA-controlled groups included in our study, as well as content posted to the 366 IRA-controlled Facebook groups not included in our analysis (as detailed above). 
Discussions with both Facebook representatives and Congressman Adam Schiff's office for the purposes of fact-checking have confirmed that this interpretation is reasonable, and there are ongoing efforts to further clarify.

\noindent\textbf{Twitter} Schiff reports finding the following content for the IRA's Facebook campaign:

\begin{itemize}
    \item More than $36,000$ Russian-linked bot accounts.
    \item $3,841$ Twitter accounts affiliated with the IRA.
    \item More than $130,000$ tweets by accounts linked to the IRA.
\end{itemize}

We had more difficulty verifying Twitter data because of ambiguity between the definitions of ``Russian-linked bot accounts'' and ``accounts affiliated or linked to the IRA''. Further the number $130,000$ was smaller than what was observed by other researchers \cite{clemson}. We attempted to fact-check with Representative Schiff's office and representatives from Twitter to resolve these queries but were unable to come to a resolution. In this paper, we therefore rely on data provided by \cite{clemson}, which aggregates historical tweets produced over time by the aforementioned IRA affiliated or linked accounts.

After excluding non-English tweets and restricting to the categories as described in the methodology section, we were left with $1,746,000$ tweets from $917$ accounts believed to be IRA-affiliated. Further restricting to our time-period of observation, we are left with $471,000$ from $360$ accounts.

\subsection{Campaign Size and Discussion}

In this discussion section, we seek to give a comparative analysis of the size of the IRA's operation in terms of staff, budget and content output. In his indictment of the IRA \cite{mueller_ira_indictment}, Special Counsel Robert Mueller states: 

\begin{quote}
    The ORGANIZATION [the IRA] employed hundreds of individuals for its online operations, ranging from creators of fictitious personas to technical and administrative support. The ORGANIZATION's annual budget totaled the equivalent of millions of U.S. dollars.
\end{quote}

As a rough comparison,  the \textit{New York Times} newsroom, according to public information, employs about 1,300 journalists, writers, copy-editors, social media producers and others \cite{nytHeadcount}. According to the \textit{Times'} 10-K report, its operating cost in $2017$ was 1.48 billion USD \cite{nyt10k}. 

The \textit{Times} uploads between 200-300 articles, blogs, and interactives a day \cite{nytContent}.
From public browsing, it appears the \textit{Times} tweets from 5-10 accounts, about 50-100 times a day. They post on Facebook  from 5-10 accounts, posting 50-100 times a day as well. 

The \textit{Washington Post} likewise has a newsroom of more than 700 staff. According to public statements, they produce an average of 500 pieces of content a day (videos, photos, articles and interactive pieces). The last available public data for the \textit{Post} shows operating costs in the range of 1.8 billion USD, in 2012\footnote{The \textit{Post} was bought in 2013 by Jeffrey Preston Bezos, and is now held by a privately owned company, Nash Holdings LLC. \cite{wpBezosPurchase}}\cite{wp10k}.
According to a brief survey of their social media accounts, they too produce between 50-100 posts on their Facebook accounts and 50-100 tweets from their Twitter accounts a day.

\textit{Buzzfeed}, according to public reporting, has a newsroom of around 460 staff. They produced roughly 200 articles a day in 2016 \cite{buzzfeedContent}. A brief survey on social media reveals that \textit{BuzzFeed} posts to Facebook between 100-200 times a day and tweets roughly the same amount.\footnote{\textit{BuzzFeed} has always been a privately held, and does not disclose costs \cite{buzzfeedExpenses}.}

\begin{figure*}[t]
    \centering
    \begin{tabular}{cccc}
    \subfloat[][Number of Facebook ads posted over time by IRA accounts.]{\includegraphics[width=.23\textwidth]{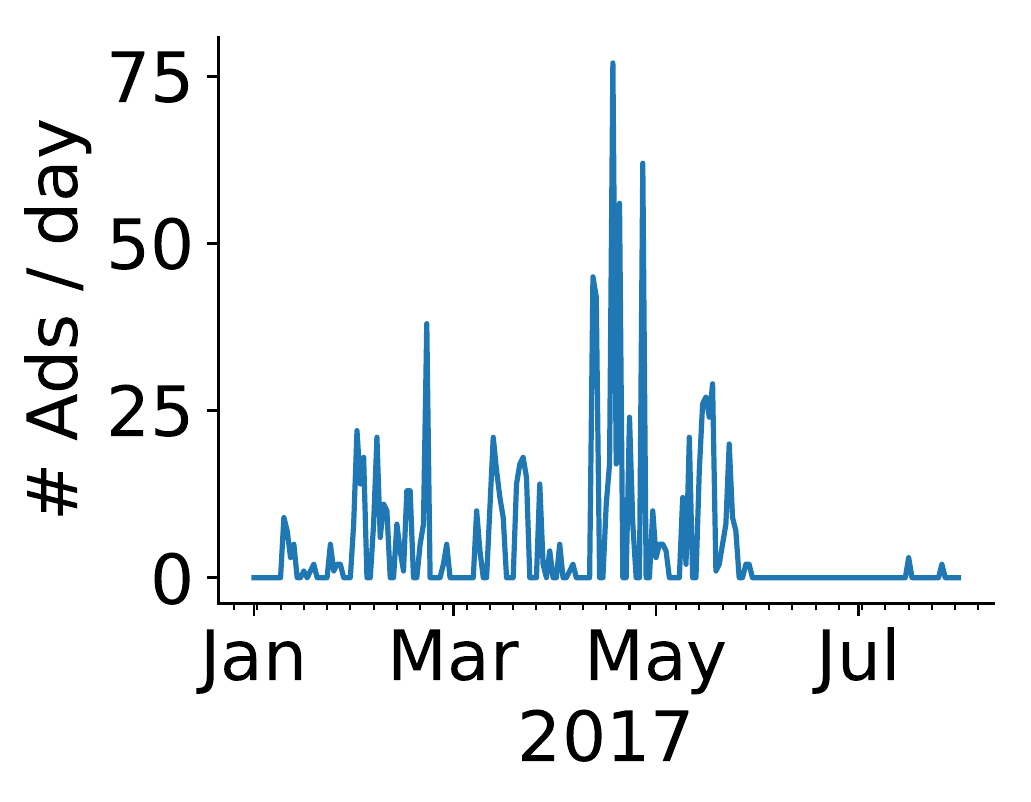}} &
    \subfloat[][Number of tweets tweeted over time by IRA handles.]{\includegraphics[width=.23\textwidth]{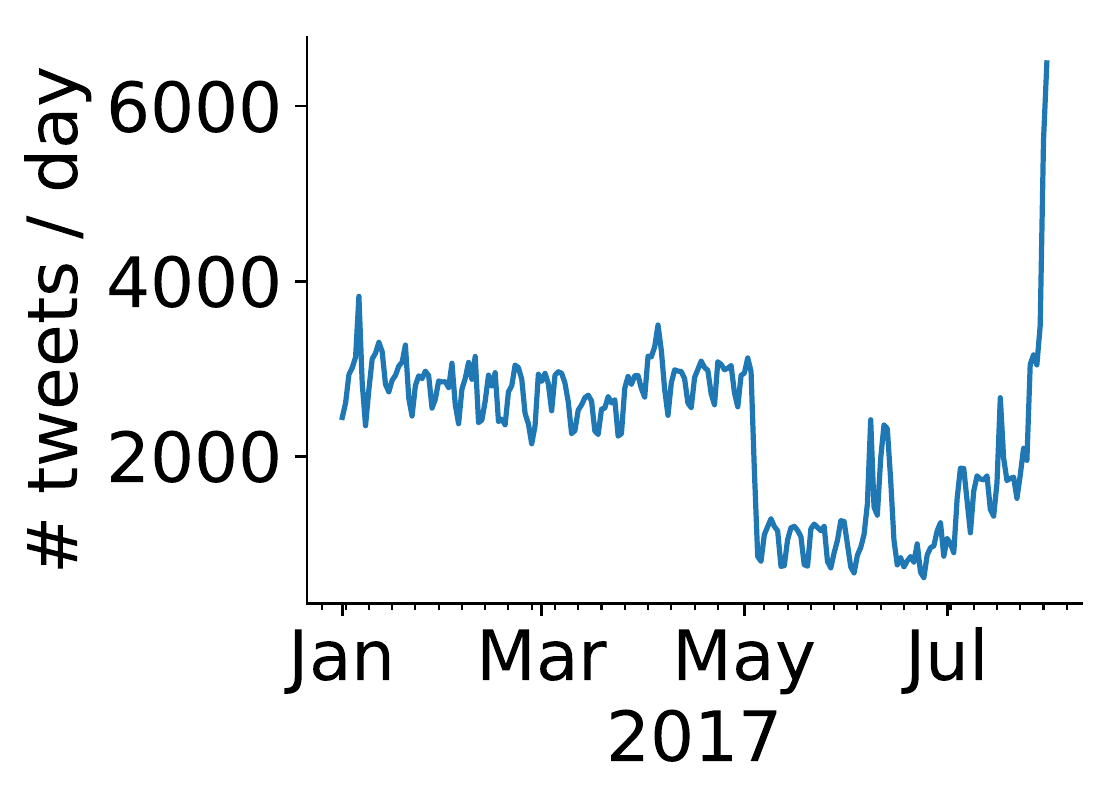}} &
    \subfloat[][Photos posted on Facebook over time to IRA accounts. 
    ]{\includegraphics[width=.23\textwidth]{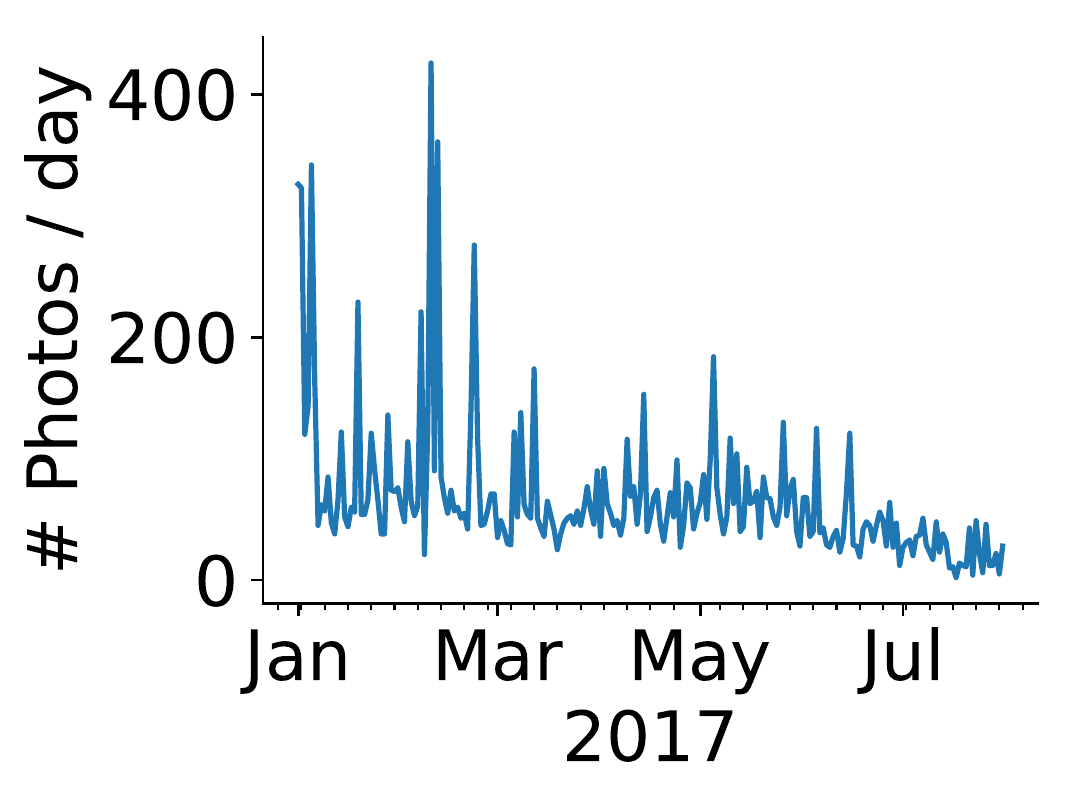}} &
    \subfloat[][Videos posted on Facebook over time to IRA accounts. 
    ]{\includegraphics[width=.23\textwidth]{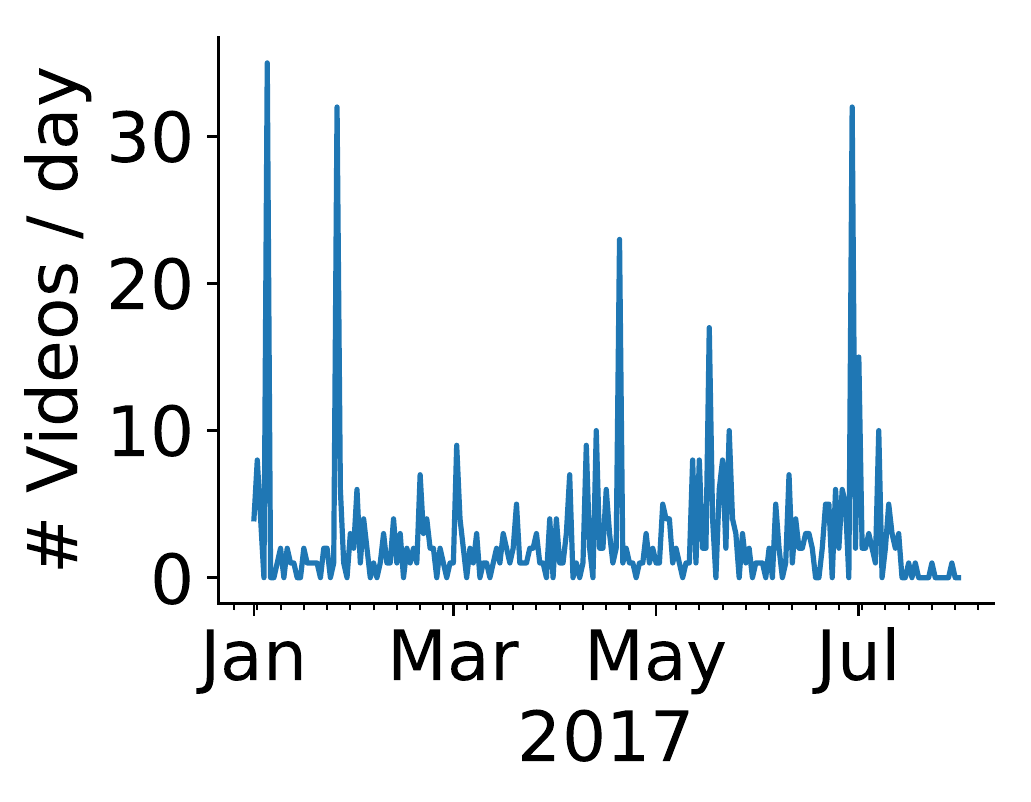}}
    \end{tabular}
    \caption{Metrics capturing the volume of unpaid photos and videos uploaded to IRA Facebook groups.}
    \label{fig:IRAActivityOverTime}
\end{figure*}

Based on material revealed from congressional testimony, the IRA generated over  $2.9$ million tweets from over $2,800$ active accounts\footnote{Of the 3841 Twitter handles in the dataset, 1034 handles posted no content during the period of observation \cite{clemson}.}. During our observation period, they tweeted at least, on average, $2,000$ times a day. They fielded $3,519$ Facebook ads, roughly $5$ ads a day during our period of interest. According to Facebook photo/video URL scraping, discussed earlier, they additionally produced at least $14,285$ photos and $575$ videos over the period of interest, or roughly $70$ photos and $3$ videos a day (Figure ~\ref{fig:IRAActivityOverTime}).

This does not include content on their owned domains. The domains we have examined show abundant material being produced. We find \url{blackmattersus.com} to be particularly expansive. By performing a site-wide scrape, we find over $3,900$ articles published, and roughly $5$ per day during our observation period. 
They logged more than $3$ new meetup links a day during our observation period.

An apples-to-apples comparison based on raw volume of output is doubtlessly flawed, since we can not compare the production time each piece of content requires at the \textit{Times}, the \textit{Post} or \textit{BuzzFeed} with the time required for content produced by the IRA. Much of the IRA's content is likely derivative or simply copied: many articles on \url{blackmattersus.com} appear duplicative, and $95\%$ of the tweets recorded from accounts of interest are retweets. (Less than 1\% of tweets from major news organizations are retweets). However, even with very conservative estimates, assuming their staff spent fractions of the time per piece of content as \textit{BuzzFeed}, we add to Mueller's estimate~\cite{mueller_ira_indictment} to project that just their \textit{content-focused} staff still numbered in the low hundreds. Their budget in the ``millions of USD'' \cite{mueller_ira_indictment} means they spent far capital per piece of content than the \textit{Times} or the \textit{Post}, but we emphasize that nevertheless, this was an effort that approached the scope of a large newsroom. 

\begin{table}[t]
\centering
\footnotesize
\begin{tabular}{|l|r|}
\hline
Media Outlet Mentioning IRA & Count of articles\\
\hline
\hline
yahoo.com                     &               830 \\
iheart.com                    &               550 \\
nbcnews.com                   &               306 \\
msn.com                       &               256 \\
washingtonpost.com            &               210 \\
reuters.com                   &               210 \\
enterprise-security-today.com &               168 \\
dailymail.co.uk               &               153 \\
\hline
\end{tabular}
\caption{\textbf{IRA mentions in Western media, top outlets.} Top outlets publishing articles mentioning ``Russian Troll'', ``IRA'', ``internet-research-agency'', ``blacktivist'', ``black-matters-us'' in URLs (calculated using \cite{gdelt}).}
\label{table:westernMediaMentions_top}
\end{table}
\begin{figure}[t]
    \centering
    \includegraphics[width=.26\textwidth]{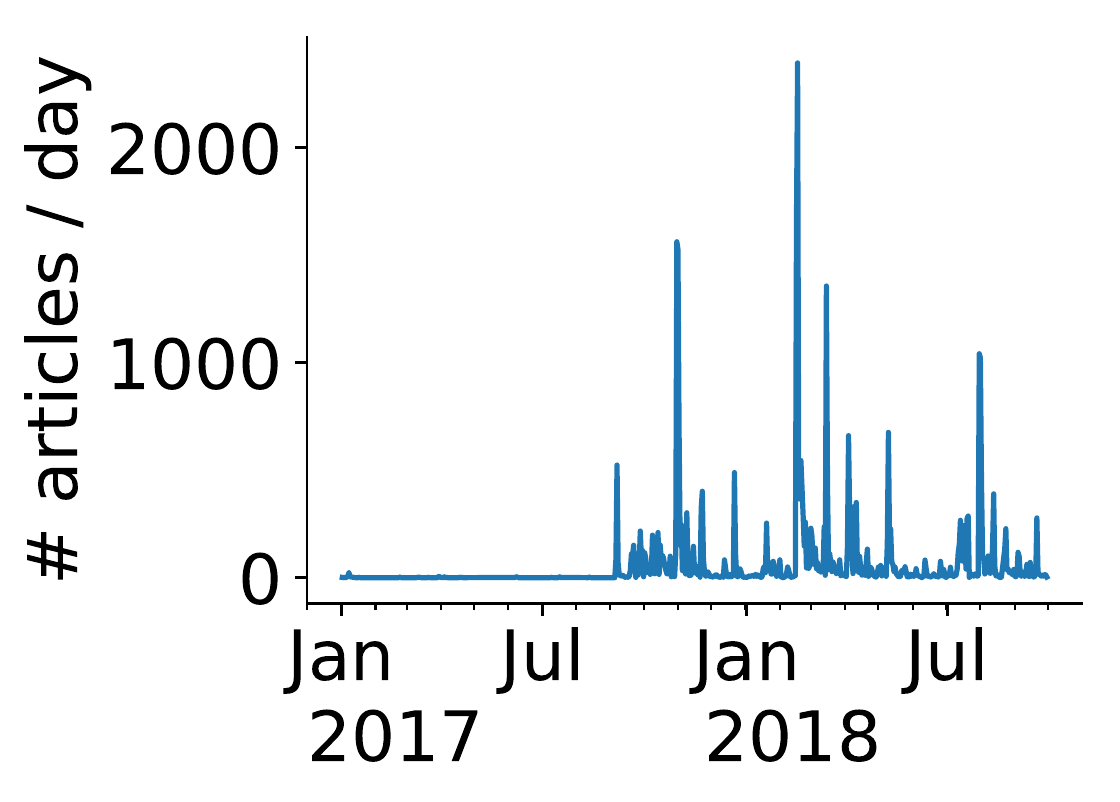}
    \caption{\textbf{IRA mentions in Western media, over time.} Number of articles in western media mentioning ``Russian Troll'', ``IRA'', ``internet-research-agency'', ``blacktivist'', ``black-matters-us'' in URLs(calculated using \cite{gdelt}).}
    \label{fig:westernMediaMentions_overTime}
\end{figure}

An important question is whether the IRA campaign was worthwhile.
As presented in Table~\ref{table:westernMediaMentions_top} and Figure~\ref{fig:westernMediaMentions_overTime}, the campaign and the agency received a significant volume of coverage from Western mainstream outlets at hundreds of articles per outlet and thousands of articles a day. These facts speak to the scale of such campaigns and the role that journalists and scientists -- present authors included -- should play in these ongoing attempts to manipulate the online information landscape.

 \section{Appendix 2: Effect of paid promotions}
As noted in the paper, the IRA spent more money on left-leaning content than on right-leaning content on Facebook during our observation period. Furthermore, traffic spikes to IRA Facebook groups were also influenced by events such as photo and video uploads in addition to paid promotions. To understand if paid promotions mattered more for some groups than for other groups, we asked: Would these groups still have had high traffic spikes had they not been advertised on Facebook?

We are limited in our ability to answer this question given that we cannot conduct a controlled study of traffic with and without paid promotions. Hence we decided to build a simple model to predict traffic to a group in the absence of a paid promotion, and use this as a comparison baseline.

We trained a model with the goal of predicting how each Facebook group would have performed \textit{had it not been promoted using a paid ad}. For this purpose we trained the model to predict the probability of a spike (a spike only using (\texttt{group url},\texttt{date}) tuples for which the url was not promoted on the given date and we omitted the promotion treatment from the input feature set. We then evaluate the predicted performance on days that articles \textit{did} receive promotions and compared our predictions. 

\begin{table}[t]
    \footnotesize
    \centering
    \begin{tabular}{|p{2.5cm}|p{5.0cm}|}
    \hline
        \textbf{Feature type (X)} & \textbf{Feature name} 
        \\
        \hline
        \hline
      historical     &  click volume (7-day window excluding the last 2 days) \\ \hline
      unpaid action  &  \#posted photos \\
                      &  \#posted videos \\\hline
      contextual     &  \#clicks on external similar urls \\
                      &  \#search queries on similar topics \\\hline
      content  &  topics present in the url (binary) \\
        \hline
    \end{tabular}
    \caption{Feature types used in counterfactual modeling.}
    \label{table:features}
\end{table}

We note that our causal analysis here is restricted. The question we ask is ``Would these groups still have had high traffic spikes had they not been advertised on Facebook?'' This is different from the question: ``What is the impact of a promotion?''. In fact, the question we ask is closer to: ``When did the promotions not matter for traffic spikes?''

\noindent {\bfseries Data and Methodology.} For this case study, we consider a specific set of Facebook groups that received both large amounts of traffic and many promotions during the train and test periods. These constraints yield six groups\footnote{These 6 groups together account for 64\% of traffic and 36\% of ads ran (15\% of ad spend) over the observation period.}. We limit our study to these groups since it is difficult to make any statements on the effect of promotions for groups that did not receive many promotions. 

We formatted our dataset as
$(\texttt{group url}, \texttt{date})$ pairs.
We trained separate $\ell_2$-constrained Logistic Regression models for each group on all $(\texttt{group url}, \texttt{date})$ pairs with \textit{no} promotions to predict the likelihood of a traffic spike to that group on the given date. Here, a ``spike" is defined as click-counts larger than the 50th percentile of all daily click-counts for the Facebook group being modeled. 
We train only on dates \emph{without} promotions, to predict the counterfactual traffic outcome on dates \emph{with} promotions. Our training data came from 1/1 to 5/1 and our test data was that collected from 5/2 to 8/1.

Table~\ref{table:features} summarizes all features used for these models. \textit{Unpaid action} features count the number of photos and videos posted by each group on the given date. \textit{Content} features are binary indicators marking topic-relevance based on our crowdsourcing task. \textit{Contextual} features track traffic to non-IRA news websites properties and volumes of search queries that were topically relevant to any of the topics of any of the groups being studied.  These features capture the possible influence of external news factors.

\begin{table}[t]
    \footnotesize
    \centering
\begin{tabular}{|l|r|r|}
\hline
\textbf{Facebook Group}  & \textbf{(a) AUC  \textit{no prom.}} &  \textbf{(b) AUC  \textit{prom.}} \\
\hline
\hline
blacktivists         &  0.965 &  0.698 \\
patriototus          &  0.876 &  0.703 \\
brownunitedfront     &  0.961 &  0.792 \\
williamsandkalvin    &  0.808 &  0.771 \\
secured.borders      &  0.792 &  0.760 \\
stopallinvaders      &   0.712 &  0.724 \\
\hline
\end{tabular}
\caption{Area under ROC curve (AUC) of prediction models tested on time-window holdout data on days (a) without promotions or (b) with promotions, ordered by the difference between the two columns.
}
\label{table:lift_model_accuracies}
\end{table}

\begin{figure}[t]
    \centering
    \includegraphics[trim={3cm 3cm 3cm 3cm},clip,width=.4\textwidth,height=.15\textheight]{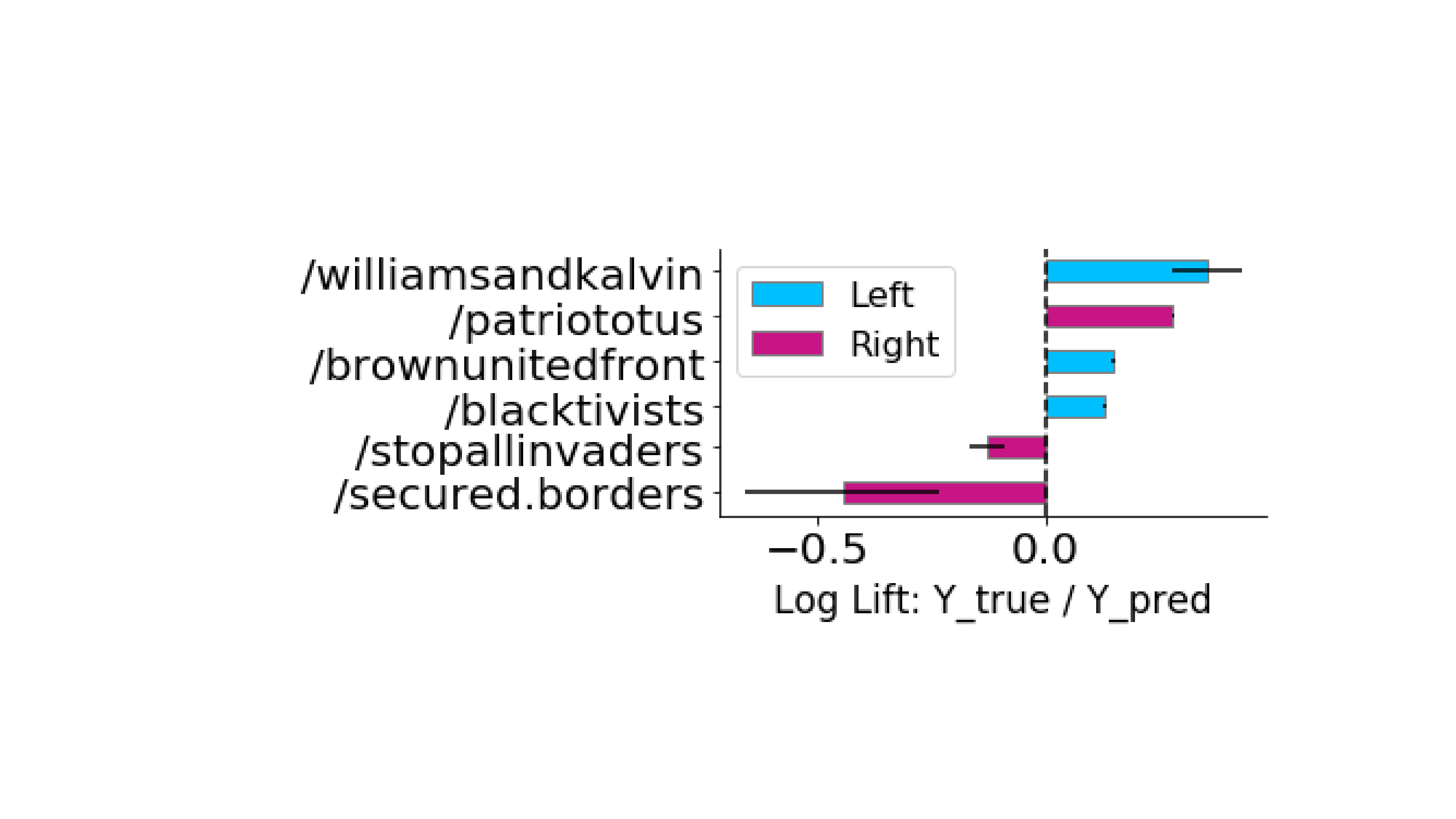} 
    \caption{\textbf{Effect of Promotions.} The average log lift of click percentile for each group on days with a paid promotion, compared the probability we predict a spike would have occurred without promotion. Positive values indicate higher importance of promotions with respect to traffic spikes.}
    \label{Fig::PromoEffectsAppendix}
\end{figure}

Additionally, instead of using traffic immediately preceding each date to calculate historical features, a buffer of 2 days is used. We purposely exclude a feature indicating whether the property had been promoted during the given day. And we took steps to ensure that no other model features would leak this information (e.g. the click volume feature only counts clicks up to two days before the spike date). We experimented with adding a feature counting past-promotions, but found little change.

We interrogated the independence relationships in our observed data to test whether our assumptions hold. We test for conditional independence between treatment and all other variables, given outcome (using $corr(x_i, x_j | y) = 0$ for all values of $y \in \{0, 1\}$, and for all $x_i$, $x_j$ pairs). We observe independence between all treatment and outcome variables ($correlation>.1, p<.1$). (We apply significance testing with Bonferroni corrections to critique our \textit{faithfulness} assumption.)\footnote{Note: we continued to see correlations between individual features in $X$, but according to the \textit{Causal Markov} assumption, this does not impact our causal claims regarding the promotion's effect on the outcome.}

These independence observations give us confidence that our assumptions are being met.

The modeling decisions we made improve the adherence of our data to the causal structure we infer. 
\begin{enumerate}
    \item Treatment assignment: We do not find strong evidence for a dependence relationship between treatment assignment and outcome, mediated through confounding factors. Based on a Causal Markov assumption, this implies that we can safely model using these features as independent \cite{Scheines97anintroduction}.
    \item Historical treatments: We also did not find significant relationships between historical promotions and current promotions, strengthening our belief that independence across different dates and urls is maintained.
\end{enumerate}

Additionally, we took steps to limit overfitting. As mentioned above, we model each Facebook Group separately. We tuned each model's $\ell_2$ hyperparameter using holdout validation data during training. We also iterated to include external news and traffic features. We choose to measure the Area Under the Receiver Operating Characteristic Curve (AUC) as a performance measure in order to handle class imbalance and also avoid picking a probability decision threshold for when a prediction should be leaning towards a spike. 

\noindent {\bfseries Findings.} The most important features across our models, in terms of average coefficient values across all models, were \textit{historical clicks}, \textit{\# posted photos}, \textit{\# posted videos}. This suggests a role for \textit{unpaid actions} in the IRA strategy.

Table~\ref{table:lift_model_accuracies} shows the AUCs for each group calculated for days with more than one promotion running (\textit{promo.} column) or with no running promotion (\textit{no prom.} column) in the test period. Again, since the models were trained only on \textit{no prom.} days in the training period, the \textit{no prom.} column is a validation test to empirically check how well the model generalizes to previously unseen data with no promotions, while the \textit{prom.} column tests our counterfactual extrapolation.

For groups with a \textit{prom.} AUC closer to a \textit{no-prom.} AUC (e.g. ``StopAllInvaders''), promotional signal is not necessary for the model to predict a spike: i.e., a variable capturing promotion-information would not have added signal and other factors, encoded in the model features, had a higher impact.
For cases where the AUC score is divergent (e.g. ``Blacktivists''), model-features were not sufficient, meaning that other factors unknown to this model, including promotions, played a significant role. Again, the reason why we look at the difference between these two columns and not only at the absolute value of each, is to decouple the ability of the model to generalize to unseen no promotion data from the ability of the model to extrapolate traffic spike predictions on dates with promotions.

Figure~\ref{Fig::PromoEffects} shows another view of these results. In Figure~\ref{Fig::PromoEffects}, we consider only the days \textit{with} promotions, and compare the observed traffic patterns (showing \textit{prom.} traffic) with our model predictions (modeling \textit{no prom.}, or the counterfactual). We calculate the logarithmic ratio between the observed traffic (in percentile) and the predicted probability of a spike on that date, as assigned by the model.

\end{document}